\documentclass[prd,12pt]{article}

\usepackage{amsmath,amssymb,graphicx,multirow,xspace,slashed}
\usepackage[colorlinks=true,urlcolor=blue,anchorcolor=blue,citecolor=blue,filecolor=blue,linkcolor=blue,menucolor=blue,pagecolor=blue]{hyperref}
\usepackage[compress,numbers]{natbib}
\usepackage{subfigure, placeins}
\usepackage{booktabs}
\usepackage{blindtext, rotating}
\usepackage{afterpage}
\usepackage{enumitem}
\usepackage{ marvosym }
\usepackage{authblk}
\usepackage{verbatim}
\usepackage{soul} 
\usepackage[normalem]{ulem}

\usepackage{floatrow}
\newfloatcommand{capbtabbox}{table}[][\FBwidth]

\usepackage[font=footnotesize,labelfont=bf]{caption}

\usepackage{lineno}

\allowdisplaybreaks

\long\def\symbolfootnote[#1]#2{\begingroup%
\def\thefootnote{\fnsymbol{footnote}}\footnote[#1]{#2}\endgroup}

\newcommand{\newc}{\newcommand}
\newc{\gsim}{\lower.7ex\hbox{$\;\stackrel{\textstyle>}{\sim}\;$}}
\newc{\lsim}{\lower.7ex\hbox{$\;\stackrel{\textstyle<}{\sim}\;$}}
\newc{\gev}{\,{\rm GeV}}
\newc{\mev}{\,{\rm MeV}}
\newc{\ev}{\,{\rm eV}}
\newc{\kev}{\,{\rm keV}}
\newc{\tev}{\,{\rm TeV}}

\newc{\mz}{M_Z}
\newc{\mpl}{M_*}
\newc{\mw}{m_{\rm weak}}
\newc{\nr}[1]{N^c_R{}_{#1}}

\def\beq{\begin{equation}}
\def\eeq{\end{equation}}
\newcommand{\bea}{\begin{eqnarray}\begin{aligned}}
\newcommand{\eea}{\end{aligned}\end{eqnarray}}
\def\bitem{\begin{itemize}}
\def\eitem{\end{itemize}}
\newcommand{\CO}{O}

 \numberwithin{equation}{section}

\newcommand\fverb{\setbox\fverbbox=\hbox\bgroup\verb}

\newbox\fverbbox

\usepackage[usenames,dvipsnames]{xcolor}

\begin{document}

\baselineskip 0.6cm

\begin{titlepage}

\thispagestyle{empty}

\begin{center}

\vskip 1cm

{\Large\bf  Precision Corrections to Fine Tuning in SUSY}
\vskip 1.0cm
{\large Matthew~R.~Buckley, Angelo Monteux and David Shih }
\vskip 1.0cm
{\it NHETC, Dept.~of Physics and Astronomy\\ Rutgers, The State University of NJ \\ Piscataway, NJ 08854 USA} \\
\vskip 2.0cm

\end{center}

\begin{abstract}

Requiring that the contributions of supersymmetric particles to the Higgs mass are not highly tuned places upper limits on the masses of superpartners -- in particular the higgsino, stop, and gluino. We revisit the details of the tuning calculation and introduce a number of improvements, including RGE resummation, two-loop effects, a proper treatment of UV vs.\ IR masses, and threshold corrections. This improved calculation more accurately connects the tuning measure with the physical masses of the superpartners at LHC-accessible energies. After these refinements,  the tuning bound on the stop is now also sensitive to the masses of the 1st and 2nd generation squarks, which limits how far these can be decoupled in Effective SUSY scenarios. We find that, for a fixed level of tuning, our bounds can allow for heavier gluinos and stops than previously considered. Despite this, the natural region of supersymmetry is under pressure from the LHC constraints, with high messenger scales particularly disfavored.

\end{abstract}

\end{titlepage}

\setcounter{page}{1}

\section{Introduction and Summary}

The naturalness of the weak scale has long been one of the best motivations for beyond-the-Standard-Model physics  at the TeV scale, with supersymmetry (SUSY) being among the most promising candidates
(for a review and original references, see e.g.~\cite{Martin:1997ns}). Given the continued non-observation of superpartners, the supersymmetric cancellation protecting the weak scale cannot be perfect, with heavier SUSY particles implying more fine-tuning of the weak scale. If we define a measure of fine-tuning $\Delta$, and require that this tuning be less than some fixed amount, then we can derive upper limits on the superpartner masses \cite{Barbieri:1987fn}.

In this paper, we will revisit the naturalness bounds on the gluino and stop masses, which (together with the Higgsino mass) are among the most important parameters for both fine-tuning and collider phenomenology \cite{Dimopoulos:1995mi,Cohen:1996vb}. We will apply a number of precision corrections to the standard SUSY tuning calculation, and we will show that they can make both a quantitative and a qualitative difference to the tuning limits on the superpartner masses. The essential point here is that the tuning measure is calculated with respect to UV parameters, defined at the messenger scale $\Lambda$ where SUSY breaking is communicated to the Standard Model superpartners, while the LHC is sensitive to the physical masses, defined in the IR at the weak scale. The two are related through the RGEs and through finite threshold corrections, and together these can have sizable effects on the naturalness bounds.

For the tuning calculation, we will use the Barbieri-Giudice measure \cite{Barbieri:1987fn}, reformulated in terms of $m_h^2\approx(125~{\rm GeV})^2$ instead of $m_Z^2$ \cite{Kitano:2006gv}, in order to better take into account the radiative corrections to the Higgs quartic: 
\beq\label{tuningmeasure}
\Delta_{M^2} = \left|{\partial \log m_h^2\over \partial \log M^2}\right|
\eeq
where $M^2$ is a UV mass-squared parameter (e.g.~$\mu^2$, $M_3^2$, or $m_{Q_3}^2$). When multiple sources of tuning are present, we take the maximum tuning as our  measure, $\Delta=\max_{\{M_i^2\}}\Delta_{M_i^2}$. Note that there is an inherent ambiguity in the definition of the measure. In particular, it is not reparametrization invariant. But while the choice of tuning measure has some arbitrariness to it, once it is decided upon, one should attempt to compute it precisely. 

Motivated by the increasingly SM-like Higgs coupling measurements \cite{ATLAStwikiHIGGS,CMStwikiHIGGS}, we will work in the decoupling limit of the Higgs sector:
\beq
V(H) = m_H^2 |H|^2 + \lambda |H|^4
\eeq
 In that case, $m_h^2= -2m_{H}^2$  and (\ref{tuningmeasure}) becomes
\beq\label{tuningmeasure1}
\Delta_{M^2} \approx  \left|{2M^2\over m_h^2}{\partial  m_{H}^2\over \partial M^2}\right| 
\eeq

For the calculation of $m_{H}^2$, it has been conventional in much of the literature
to work in the leading-log (LL) approximation (see however \cite{Essig:2007kh,Arvanitaki:2013yja,Casas:2014eca}  for notable exceptions). In this approximation,   the  quadratic sensitivity of the Higgs mass-squared parameter to the higgsino, stop and gluino soft masses arises at tree level, one-loop and two-loops respectively: 
\begin{itemize}

\item{} Higgsinos:
\beq\label{deltamhusqhiggsino}
\delta m_{H}^2 = |\mu|^2
\eeq
\item{} Stops:\footnote{In this paper, we will be neglecting the $A$-terms (i.e.\ we are assuming they are small). We will also treat $m_{Q_3}^2$ and $m_{U_3}^2$ as separate UV parameters for the purposes of the tuning computation. In some UV completions, such as gauge mediation, they may in fact be correlated or even equal. This would strengthen the tuning bounds on stops by up to a factor of $\sqrt{2}$ relative to what will be quoted in this work.}
\beq\label{deltamhusqstop}
\delta m_{H}^2\sim -{3\over8\pi^2}y_t^2 (m_{Q_3}^2+m_{U_3}^2) \log {\Lambda\over Q}
\eeq
\item{ } Gluinos:\footnote{Notice that our formula corrects a factor of 2 mistake in \cite{Papucci:2011wy}. This correction alone relaxes their gluino tuning bounds by a factor of $\sqrt{2}$, which is numerically quite significant. }
\beq\label{deltamhusqgluino}
\delta m_{H}^2\sim -{g_3^2y_t^2\over 4\pi^4}|M_3|^2\left(\log{\Lambda\over Q}\right)^2 
\eeq
\end{itemize}
Here $\Lambda$ is the messenger scale of SUSY breaking, and $Q$ is the IR scale, conventionally taken to be 1~TeV in many works (see e.g.~\cite{Papucci:2011wy,Casas:2014eca}). For definiteness, we are assuming here (and throughout this work) that the stops and gluinos contribute as in the MSSM.

The higgsino formula is fairly accurate as is. For better-than-10\% tuning ($\Delta\le10$), we need 
\beq\label{mhiggsinobound}
\mu\lesssim 300~\mbox{GeV}.
\eeq
Meanwhile, the stop and gluino formulas are rather imprecise, and the purpose of this paper is to include a number of higher order corrections.  We identify five such corrections in this paper:
\begin{enumerate}

\item While it is common in the literature to use the LL approximation for gluinos, the next-to-leading-log (NLL) correction can be numerically important at lower messenger scales:
\beq\label{deltamhusqgluinonll}
\delta m_{H}^2\sim -{g_3^2y_t^2\over 4\pi^4}|M_3|^2\left(\log{\Lambda\over Q}\right)^2 -{g_3^2y_t^2\over 4\pi^4}|M_3|^2\log{\Lambda\over Q}.
\eeq

\item The LL formulas refer to $y_t$ and $\alpha_s$, but these run considerably with the RG. In particular, for moderate to large $\tan\beta$ in the MSSM, $\alpha_s$ and $y_t$ both decrease considerably in the UV. If one uses $y_t$ and $\alpha_s$ defined at the weak scale, one can considerably overestimate the tuning. This effect tends to be more important at higher messenger scales.

\item Even more importantly, in the LL approximation, there is no difference between $m_{Q_3}^2$, $m_{U_3}^2$, $M_{3}$ evaluated in the UV and in the IR. In reality, these masses evolve quite a bit with the renormalization group. In fact, as we will see below, it is often the case that the IR masses are considerably larger than the UV masses. This can further relax the tuning bounds. 

\item The energy $Q$ is an IR renormalization group scale. 
A proper treatment includes threshold corrections to $m_{H_u}^2$ that would remove the $Q$ dependence, effectively replacing it with a physical scale, for example $m_{\rm stop}$ or $M_{\rm gluino}$. 

\item  Finally, the gluino and stop masses are subject to their own threshold corrections, leading to a difference between the running IR masses and the pole masses.

\end{enumerate}

In Section~\ref{sec:tuning},
we will address these issues in turn, using the fully integrated RGEs (called the ``transfer matrix" in \cite{Knapen:2015qba}) for items 1-3, the two-loop effective potential \cite{Martin:2002iu} for item 4, and the one- and two-loop pole mass formulas from \cite{Pierce:1996zz} and \cite{Agashe:1998zz} respectively for item 5. Some of these corrections, having to do with RG effects, were previously studied in \cite{Essig:2007kh,Casas:2014eca}. We will expand on these results, in particular adding in important finite threshold corrections to $m_{H}^2$ and the stop and gluino masses.

In a companion paper \cite{Buckley:2016kvr}, we have reinterpreted the latest LHC searches post-ICHEP and used them to understand the current experimental constraints on the natural SUSY parameter space, as determined by to the improved fine-tuning calculation described in this work. We considered a set of simplified models for natural SUSY, starting from the most ``vanilla" case (the MSSM with $R$-parity conservation and flavor-degenerate sfermions), and then proceeding to more complicated scenarios which can better hide SUSY at the LHC. Because of the large valence squark cross sections, we found that a good strategy to relax LHC limits is to decouple the first and second generation of squarks (``Effective SUSY'') and only keep light the squarks  which are most important for the fine-tuning, namely $\tilde t_L, \tilde b_L,\tilde t_R$ \cite{Dimopoulos:1995mi,Cohen:1996vb}.\footnote{We also note that in order to reduce squark production rates, one could also decouple only the squarks of the first generation \cite{Mahbubani:2012qq}. In this case, alignment between 1st and 2nd generation squarks would be needed to avoid large contributions to kaon mixing, but on the other hand the RGE contributions lowering the IR stop mass would be reduced by half. Up and down squarks could then be a factor of $\sim\sqrt2$ heavier and give the same tuning bounds on the stop mass discussed here. } However, as will be shown here, this does not come without a cost. Through the threshold corrections and the RG running, heavy 1st/2nd generation squarks improve the tuning on gluinos, but worsen the tuning for stops by a larger degree. That is to say, as the 1st/2nd generation squark masses are increased, a given amount of tuning allows for heavier gluinos but requires lighter stops.
Given this tension between gluino and stop tuning, and taking into account the LHC constraints   \cite{Buckley:2016kvr}, we find that a ``sweet spot" for 1st/2nd generation squark masses is in the $\sim 2$--5~TeV range. 

 In Section~\ref{sec:results}, we combine all of the precision corrections to tuning and present the fully-natural regions (taken here to mean $\Delta\le 10$) in the gluino-stop mass plane, as a function of the mass of  the 1st/2nd generation squarks and the messenger scale. 
Because of the effects discussed, namely the threshold corrections to gluinos from stops and the large gluino contribution to the RGE for the stop mass, we find that natural regions are shaped like wedges with heavy gluinos favoring heavy squarks and vice versa (see Figs.~\ref{fig:wedge} and \ref{fig:tuning_all} for examples of this behavior). For very low messenger scales, $\Lambda=20$~TeV, we find that fully-natural gluinos (stops) should be below 2.2 (1.5)~TeV, with some dependence on the 1st/2nd generation squark masses. 
For larger messenger scales, the fully-natural region shrinks, with $(m_{\tilde g},m_{\tilde t})<(1.5,1.2)~$TeV for $\Lambda=100$~TeV. This is close to the current LHC limits even in the best possible scenarios, e.g. with decoupled squarks and RPV decays of the higgsino \cite{Buckley:2016kvr}. In the light of this, we conclude that fully-natural supersymmetry requires messengers at or below 100~TeV, likely with 1st and 2nd generation squarks significantly heavier than the stops. (On the other hand, percent-level-tuned SUSY is considerably less constrained, with much higher messenger scales still allowed.) 

We should note here that for $\Delta\sim 10$ to be meaningful,  there should be an additional contribution coming from a sector beyond the MSSM in order to raise $m_h$ to 125~GeV. This could be e.g.\ the NMSSM  (see \cite{Ellwanger:2009dp} for a review and original references) or non-decoupling $D$-terms \cite{Batra:2003nj,Maloney:2004rc}. Otherwise, as is well known  \cite{Hall:2011aa, Heinemeyer:2011aa, Arbey:2011ab, Arbey:2011aa, Draper:2011aa, Carena:2011aa}, the  125 GeV Higgs in the MSSM requires either multi-TeV $A$-terms or $\gtrsim$~10~TeV stops; at best the resulting fine-tuning is a few percent (for a recent discussion, see e.g.~\cite{Casas:2014eca}). Our implicit assumption here is that this additional sector is such that it does not modify the calculation of the tuning with respect to the stops and gluinos. In other words, we assume that
\beq
m_H^2 = m_{H_u}^2 + |\mu|^2 + \Delta m_H^2
\eeq
where $m_{H_u}^2$ is as in the MSSM, and $\Delta m_H^2$ is the additional contribution that depends at most weakly on the stop and gluino masses. If these assumptions do not hold, then the tuning calculation should be revisited. Nevertheless, we expect the effects we have considered in this paper would need to be taken into account in any supersymmetric model, more complete or otherwise. So at the very least, the treatment here should be taken as a template for future works. 

Finally, in Section~\ref{sec:conclusion}, we conclude with a brief summary of our results, together with a discussion of some well-motivated directions for future model-building. These include ways to achieve an effective SUSY spectrum with an ultra-low messenger scale, as well as loopholes to the conventional tuning bounds -- models beyond the MSSM (e.g.~Dirac gluinos) with reduced contributions of the higgsino, stop and/or gluino masses to the renormalization of the  weak scale.

\section{Precision corrections to fine-tuning \label{sec:tuning}}

\subsection{Transfer matrix RGEs: $m_{H_u}^2$}
\label{sec:xfermatmhu2}

As discussed in the Introduction, the leading-log tuning formulas (\ref{deltamhusqstop})-(\ref{deltamhusqgluino}) have a number of practical drawbacks. They neglect higher-order terms, they refer to couplings and soft parameters at an indeterminate scale, and they refer to an arbitrary IR renormalization group $Q$. In this subsection and the next, we will remedy the first two deficiencies (items 1 and 2 in the list above) by employing the fully-integrated two-loop RGEs (as derived from \texttt{SARAH} \cite{Staub:2013tta}) instead of their LL approximation,  and then translating the tuning bounds into ones on running IR masses  (item 3). The importance of integrating the RGEs and rephrasing the tuning bounds in terms of the IR parameters was previously emphasized in  \cite{Essig:2007kh,Arvanitaki:2013yja,Casas:2014eca}.

As is well-known, integrating the MSSM RGEs between a UV scale $\Lambda$ and an IR scale $Q$ results in a (bi)linear map -- a sort of ``transfer matrix" -- that relates the soft parameters defined at these scales. Let the dimension-one soft parameters be denoted by $M$ and let the dimension-two soft parameters be denoted by $m^2$. For each dimension-one soft parameter $M$, the transfer matrix takes the form 
\beq
M(Q) = \sum_{M'} A_{M}^{M'}(Q;\Lambda)M'(\Lambda)
\eeq
while for each dimension-two soft parameter $m^2$, the matrix takes the form
\beq
m^2(Q) = \sum_{m'^2} A_{m^2}^{m'^2}(Q;\Lambda)m'^2(\Lambda) + \sum_{M',M''}A_{m^2}^{M'M''}(Q;\Lambda)M'(\Lambda)M''(\Lambda)
\eeq
In what follows, we will generally suppress the $Q$ and $\Lambda$ dependence of the transfer matrix coefficients to avoid cluttering the equations. 

\begin{figure}[t]
\centering
\includegraphics[width=0.8\textwidth]{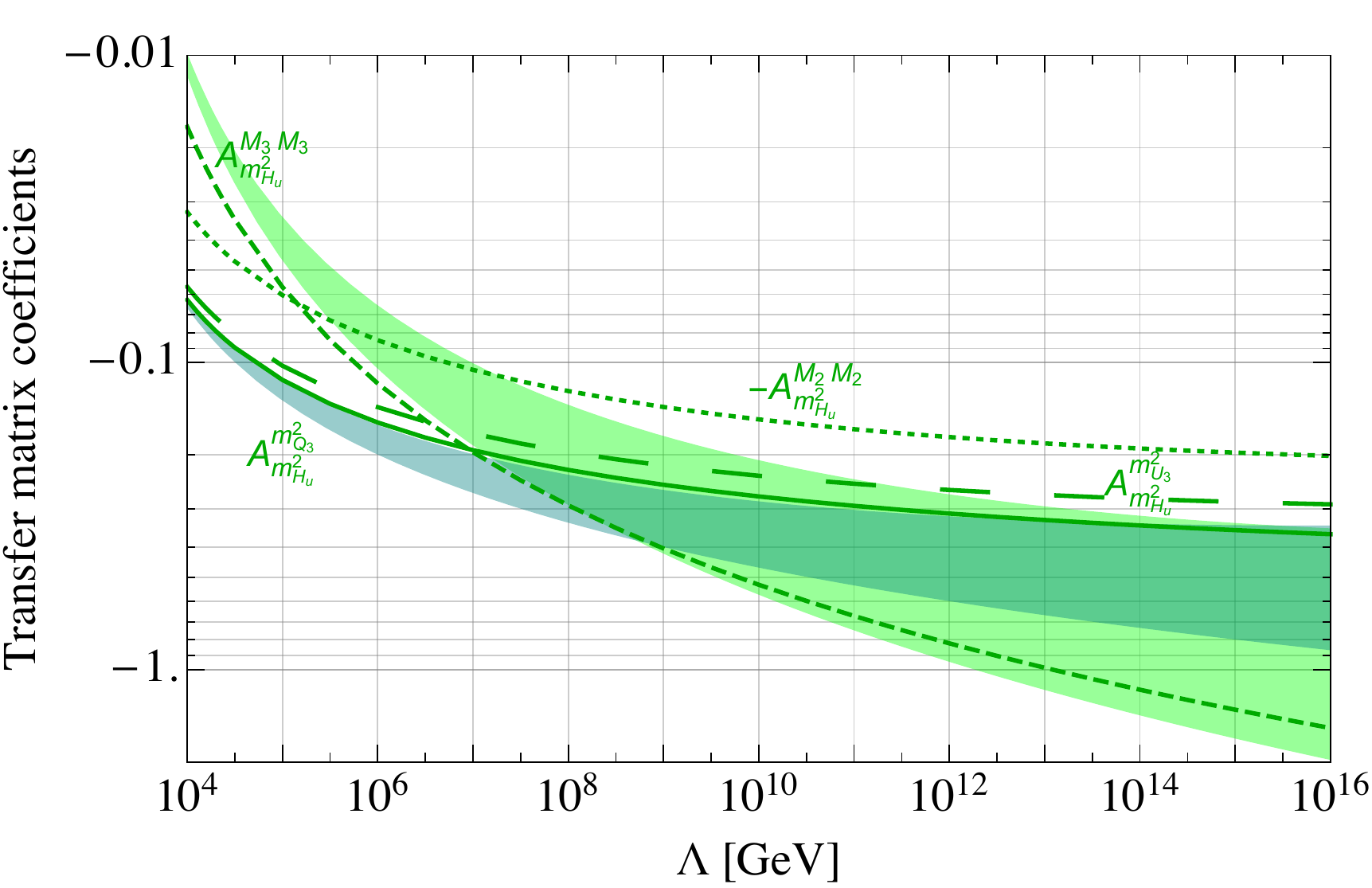} 
\caption{\small{Relevant transfer matrix coefficients for $m_{H_u}^2$ as function of the messenger scale $\Lambda$, for $Q=1$~TeV and $\tan\beta=20$. The solid and long-dashed lines show coefficients for the terms proportional to the squared squark masses $m_{Q_3}^2$ and $m_{U_3}^2$, while the short-dashed and dotted lines are the coefficient for the terms porportional to $M_3^2$ and $M_2^2$. Shaded bands show the intrinsic uncertainties in the LL approximations, by taking the LL formulas (\ref{deltamhusqstop})-(\ref{deltamhusqgluino}) with $y_t$ and $\alpha_s$ evaluated at the weak scale (lower boundary) or at the messenger scale (upper boundary). 
}}
\label{fig:xfermat_mhu2}
\end{figure}

For $m_{H_u}^2$, the dominant terms in the transfer matrix relation are given by:
\beq
\label{xfermhusq}
m_{H_u}^2(Q)=
A_{m_{H_u}^2}^{m_{Q_3}^2}\ m_{Q_3}^2(\Lambda)+A_{m_{H_u}^2}^{m_{U_3}^2}\  m_{U_3}^2(\Lambda)+A_{m_{H_u}^2}^{M_3\,M_3}\ |M_3(\Lambda)|^2+\dots.
\eeq
The masses of the other superparticles contribute less to $m_{H_u}^2$, being suppressed either by small Yukawa couplings or by $\alpha_{1,2}$ (note also that we have assumed small $A$-terms in our analysis). This is the transfer matrix upgraded version of the LL formulas   (\ref{deltamhusqstop})-(\ref{deltamhusqgluino}). Indeed, one can check analytically that expanding the coefficients $A_{m_{H_u}^2}^{m_{Q_3}^2}(Q;\Lambda)$, etc.\ in powers of $t=\log Q/\Lambda$ reproduces the leading order behavior shown in (\ref{deltamhusqstop})-(\ref{deltamhusqgluino}). Using \eqref{tuningmeasure1} and \eqref{xfermhusq},  one can compute the upper bound on a UV mass parameter $m^2(\Lambda)$ for a given fine-tuning level $\Delta$:
\beq\label{tuningUV}
|m^2(\Lambda)| < m_h^2 \times\frac{\Delta}{2 |A_{m_{H_u}^2}^{m\,m}|}
\eeq
In the following subsections, we will apply a number of corrections and successively improve this into a bound on the physical masses.

A plot of the transfer matrix coefficients, together with a comparison to the LL approximation, is shown in Fig.~\ref{fig:xfermat_mhu2}, as a function of the messenger scale $\Lambda$, for $Q=1$~TeV and $\tan\beta=20$. (Unless otherwise stated, this will be our benchmark value of $\tan\beta$ throughout the paper.) For the latter, the RG scale of the running couplings is varied from $Q$ to $\Lambda$, demonstrating one of the inherent ambiguities in the LL approximation formulas.

It can be seen from Fig.~\ref{fig:xfermat_mhu2} that the $Q_3$ and $U_3$ contributions to $m_{H_u}^2$ are similar, with differences amounting to only about 10\%  (20\%) at low (high) messenger scales. Given our small $A$-term assumption, this will correspond to similar tuning bounds on $\tilde t_L$ and $\tilde t_R$, with the bound on the latter slightly weaker than on the former. Unless explicitly specified, we will be referring to $\tilde t_L$ when showing bounds on the stop.

We also show  with a dotted line in Fig.~\ref{fig:xfermat_mhu2} the transfer matrix coefficient for $M_2^2$; it is actually not much smaller than the others. For $\Delta_{M_2^2}<10$, this gives an upper bound on the wino mass ranging from 1.5~TeV to 500~GeV for $\Lambda$ between $10$~TeV and $10^{16}$~GeV (note that $M_2$ itself does not run more than $20\%$ between the messenger scale $\Lambda$ and the IR scale $Q$, so the the bounds on UV and IR parameters are similar). While natural SUSY spectra usually focus on stops, gluinos and higgsinos, it should be noted that, given present collider constraints on gluinos and squarks \cite{Buckley:2016kvr}, a natural wino will typically participate in the cascade decays of superpartners.

\subsection{Transfer matrix RGEs: stop and gluino masses}
\label{sec:xfermatstopgluino}

Next we will address the difference between the UV soft mass (which the BG measure $\Delta$ is calculated with respect to) and the IR soft mass (which is more physically relevant, especially for the collider phenomenology). This is item \#3 on the list presented in the Introduction.  We will focus on the stop and gluino masses; for higgsinos, the running from the messenger scale is generally negligible. 
 
 For the gluinos the translation from UV to IR is straightforward. As is well-known, at one-loop, the running gaugino masses simply scale with the gauge-couplings squared:
\beq\label{xfergluino}
M_3(Q) \approx  {g_3^2(Q)\over g_3^2(\Lambda)} M_3(\Lambda).
\eeq
In other words, $A_{M_3}^{M_3}\approx  {g_3^2(Q)\over g_3^2(\Lambda)} $ with $A_{M_3}^{M'}\approx 0$ for $M'\ne M_3$. Since $g_3$ is asymptotically free in the MSSM, it is always the case that $M_3$ grows in magnitude in the IR. So the tuning bound on the IR gluino mass will always be relaxed as compared to the bound on the UV gluino mass.

\begin{figure}[t]
\centering
\includegraphics[width=0.8\textwidth]{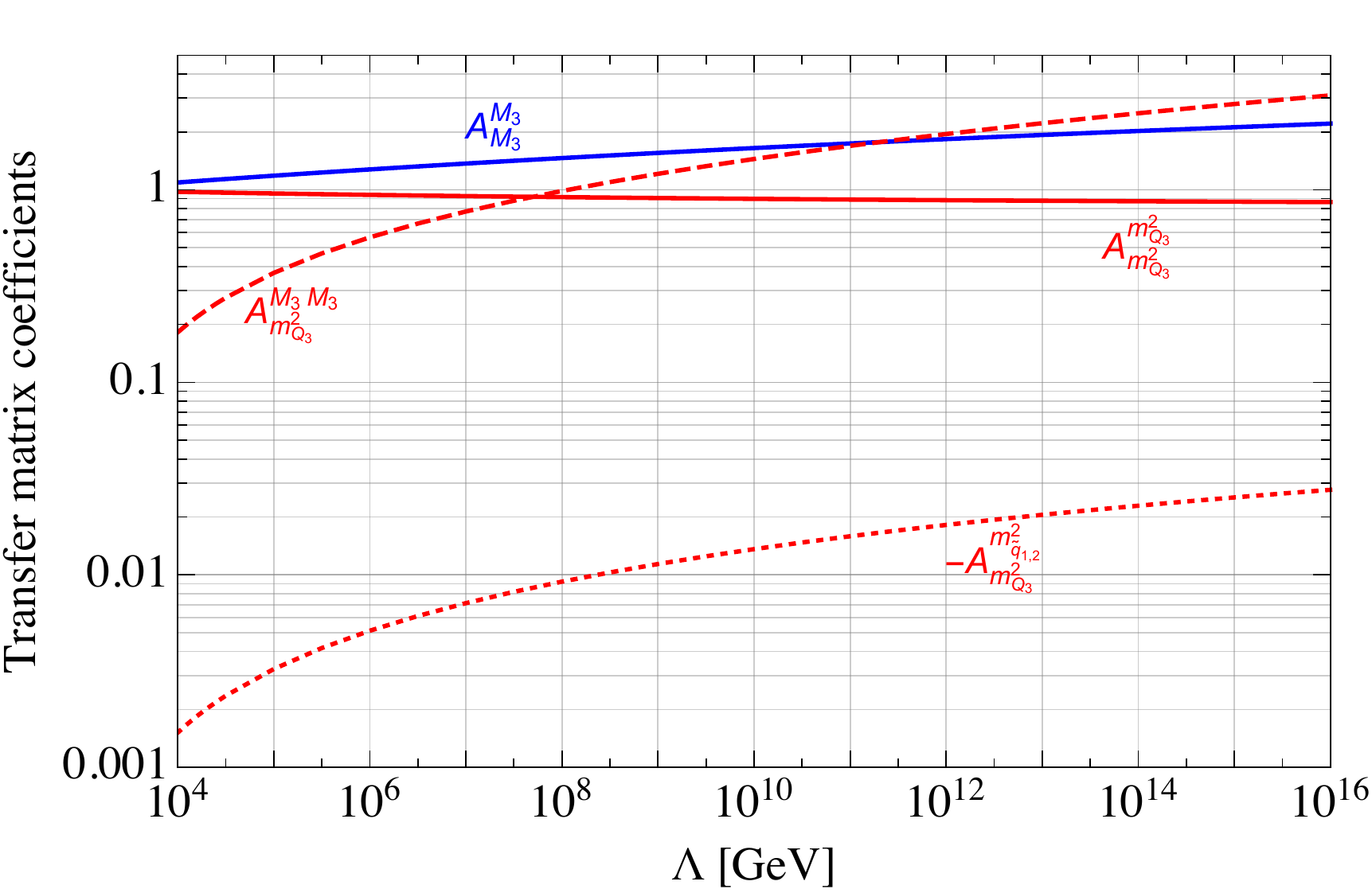} 
\caption{\small{Relevant transfer matrix coefficients for $M_3$ (solid blue) and $m_{Q_3}^2$ (red), as function of the messenger scale $\Lambda$, for $Q=1$~TeV and $\tan\beta=20$. Note that $A_{m_{Q_3}^2}^{m_{\tilde{q}_{1,2}}^2}$ is negative, unlike the other coefficients. 
}}
\label{fig:xfermat_M3Q3}
\end{figure}

For stops, the most important terms are the stop mass-squared itself and the gluino mass. The 1st/2nd generation squark squared masses contribute irreducibly at two-loop in the RGEs, proportional to $g_3^4$, and can become important  if they are much heavier.\footnote{There is also a potentially dangerous 1-loop contribution to the stop and Higgs soft mass-squareds from the hypercharge $D$-terms \cite{Dimopoulos:1995mi,Cohen:1996vb}, but these are absent if the 1st/2nd generation squark masses are decoupled in degenerate $SU(5)$ multiplets, which we assume throughout this work.} So we have
\beq\label{xferstop}
m_{Q_3}^2(Q)= A_{m_{Q_3}^2}^{m_{Q_3}^2}\ m_{Q_3}^2(\Lambda) +A_{m_{Q_3}^2}^{m_{\tilde q_{1,2}}^2}\ m_{\tilde q_{1,2}}^2(\Lambda) +A_{m_{Q_3}^2}^{M_3\,M_3}\ |M_3(\Lambda) |^2+ \dots
\eeq
and similarly for $m_{U_3}^2$. Here and below, we have taken the first and second generation squark masses to be the same,\footnote{Technically only $m_{Q_3}$ and $m_{U_3}$ (setting the masses of $\tilde t_L$, $\tilde b_L$ and  $\tilde t_R$) contribute at one-loop to $m_{H_u}^2$, so one could raise $m_{D_3}$ without directly affecting tuning. However, as with the other squarks, the right-handed sbottom  enters the stop RGEs and pulls down the IR stop mass, worsening the tuning. In this work we take $\tilde  b_R$ to be at the same scale as other third generation squarks to minimize this effect. Having all 3rd generation squarks at one scale should also be simpler from the model-building prospective.}  
\beq\label{squarkassumption}
m_{\tilde q_{1,2}}^2\equiv m_{Q_{1,2}}^2=m_{D_{1,2}}^2=m_{U_{1,2}}^2
\eeq
so that the transfer matrix coefficient is 
\beq
A_{m_{Q_3}^2}^{m_{\tilde q_{1,2}}^2}\equiv A_{m_{Q_3}^2}^{m_{Q_{1,2}}^2}+A_{m_{Q_3}^2}^{m_{U_{1,2}}^2}+A_{m_{Q_3}^2}^{m_{D_{1,2}}^2}
\eeq
Importantly, $A_{m_{Q_3}^2}^{M_3\,M_3}>0$, i.e.\ the gluino always pulls up the squark masses. This is a significant effect in the context of natural SUSY, as it can allow heavier-than-expected stops.
On the other hand, $A_{m_{Q_3}^2}^{m_{\tilde q_{1,2}}^2}<0$, that is, the first and second generations reduce the IR stop mass, worsening the fine-tuning for large hierarchies between them and the third generation squarks. For very heavy 1st/2nd generation squarks or too-high messenger scales, the stop squarks can even become tachyonic due to this effect
\cite{ArkaniHamed:1997ab,Agashe:1998zz}.

\begin{figure}[t]
\centering
\includegraphics[width=0.75\textwidth]{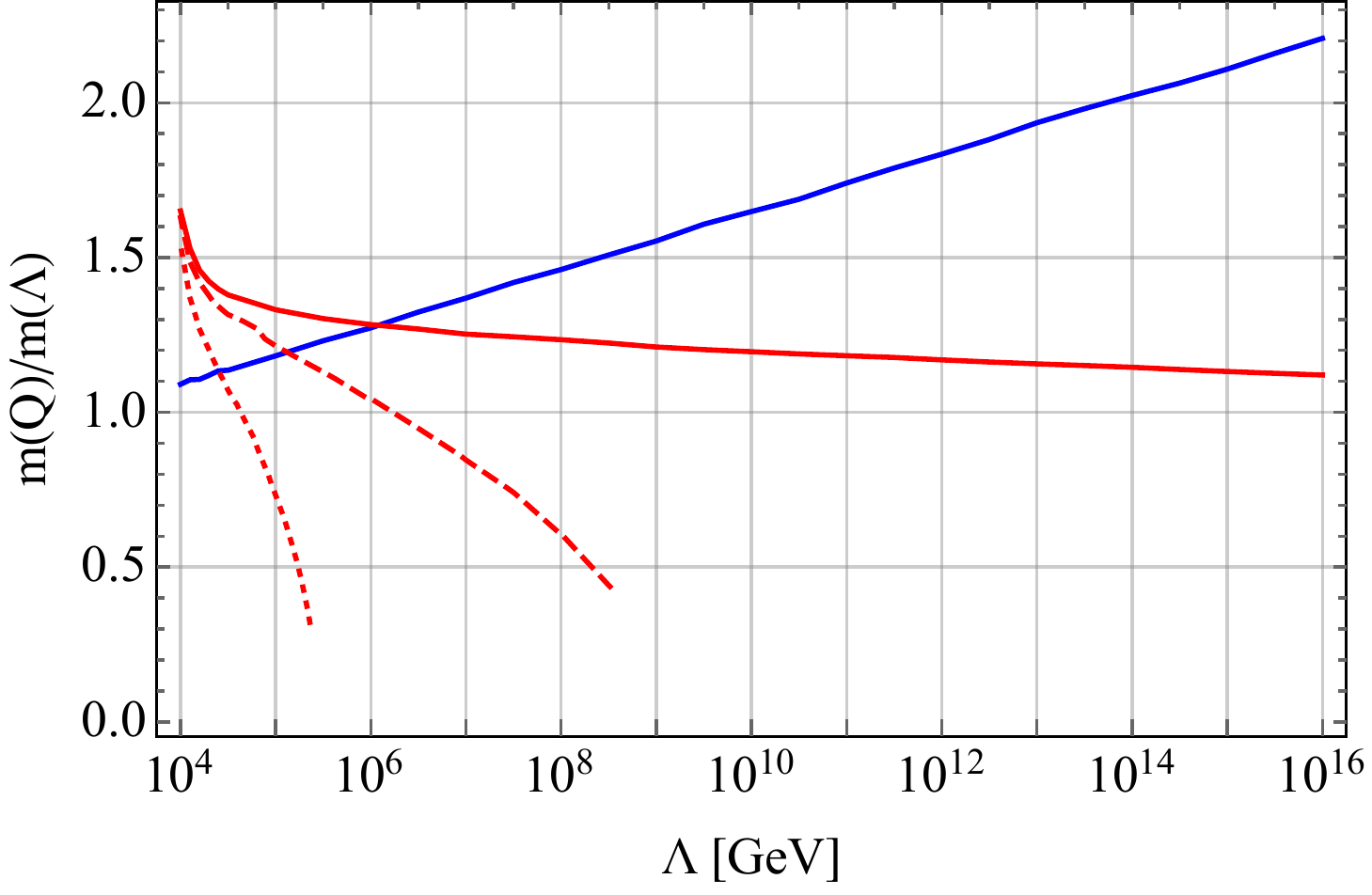} 
\caption{\small{Ratio of the 10\% tuning bound on the IR (at $Q=1$~TeV) vs. UV running mass for the gluino (blue line - this is simply $A_{M_3}^{M_3}$) and the stop (red lines), for 1st and 2nd generation squarks either degenerate with the stop (solid red) or set at 5~TeV (dashed red) or 10~TeV (dotted red). 
}}
\label{fig:IRvsUV}
\end{figure}

Shown in Fig.~\ref{fig:xfermat_M3Q3} are the transfer matrix coefficients $A_{M_3}^{M_3}$, $A_{m_{Q_3}^2}^{m_{Q_3}^2}$, $A_{m_{Q_3}^2}^{m_{\tilde q_{1,2}}^2}$, $A_{m_{Q_3}^2}^{M_3M_3}$ vs.\ the messenger scale $\Lambda$, for $Q=1$~TeV. In Fig.~{\ref{fig:IRvsUV}}, we show ratios of the IR to UV stop and gluino masses vs.\ $\Lambda$. While for the gluino there is a simple one-to-one mapping between UV and IR (given by $A_{M_3}^{M_3}$), for the IR stop mass \eqref{xferstop} one has to also specify the gluino and other squark masses. For definiteness (and in anticipation of our results in the next section), here we set gluino and stop UV masses to their $\Delta=10$ upper limits given by \eqref{tuningUV}.
We see that for the gluino, the IR mass is considerably higher than the UV mass due to the running of $g_3$. For the stops, the gluino lifts the IR mass, while the   1st/2nd generation squarks pull it down. In particular, it can be seen that highly decoupled squarks do not allow fully-natural stops with high messenger scales, an effect which would be lost if only considering UV parameters.

\subsection{Higgs potential threshold corrections}

Here we will consider the $m_{H_u}^2$ threshold corrections that remove the $Q$-dependence 
in the LL tuning formulas (item \#4 on the list in the Introduction). These can be obtained from a two-loop effective potential calculation \cite{Martin:2002iu}.  Up to two-loops, the most important terms in the effective potential $V$ are
\bea\label{effpotgen}
& V^{(0)} \supset v_u^2(|\mu|^2+m_{H_u}^2) \\
& 16\pi^2V^{(1)} \supset 6(h(m_{\tilde t_1}^2)+h(m_{\tilde t_2}^2)-2h(m_{t}^2)) \\
& (16\pi^2)^2V^{(2)} \supset 8g_3^2\Big(F_{FFS}(m_{t}^2,m_{\tilde g}^2,m_{\tilde t_i}^2)-2Re[L_{\tilde t_i}R_{\tilde t_i}^*]F_{\bar F\bar F S}(m_{t}^2,m_{\tilde g}^2,m_{\tilde t_i}^2)\Big)
\eea
Here $X=L,R$ are the stop mixing matrices, $\tilde t_X=\sum_i X_{\tilde t_i} \tilde t_i$. The function $h(x)= {x^2\over4}\left( \log(x/Q^2)-{3\over2}\right)$ was defined in \cite{Martin:2002iu} and describes the one-loop stop/top corrections. The explicit dependence on the gluino mass first enters in at two-loops; these all come from the fermion-fermion-scalar terms $F_{FFS}$ and $F_{\bar F\bar F S}$ defined in \cite{Martin:2002iu}. Additional terms will acquire $M_3$ dependence through the RG (e.g.~\ terms that involve the stop and left-handed sbottom masses-squared and the $A$-terms)  but we find these to be numerically subleading. 

\begin{figure}[t!]
\centering
\includegraphics[scale=1.2]{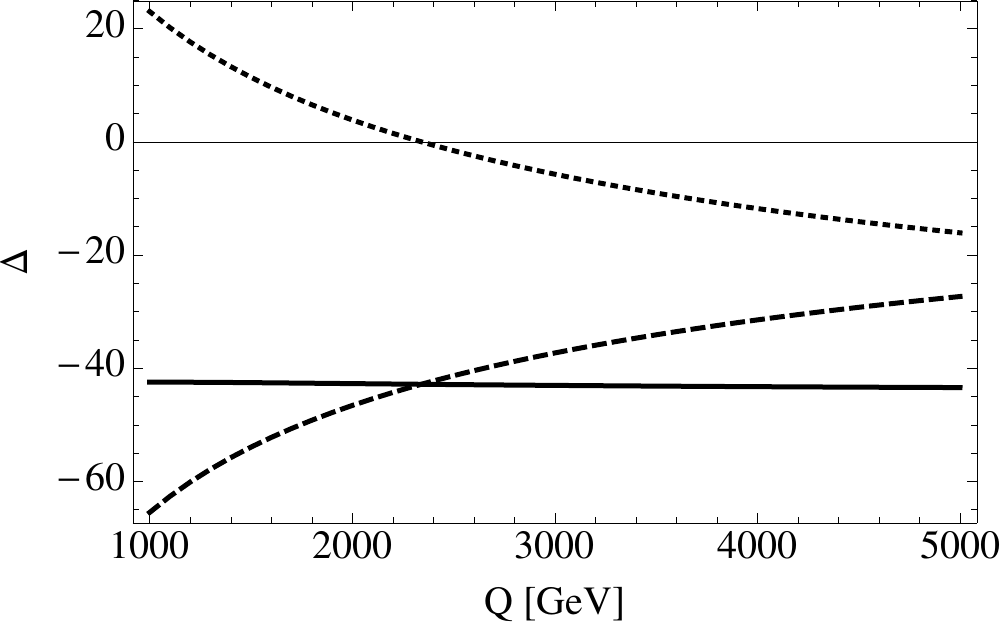}
\caption{\small{The gluino tuning vs RG scale for a benchmark point $m_{Q_3}(\Lambda)=m_{U_3}(\Lambda)=1$~TeV, $M_3(\Lambda)=3$~TeV, $\Lambda=100$~TeV, $\tan\beta=20$. The dashed curve is derived from $m_{H_u}^2(Q)$, the scale-dependent tree-level term in (\ref{effpotgen}). As can be seen, this is subject to large RG-scale uncertainty. The dotted curve shows the threshold corrections from $V^{(1)}$ and $V^{(2)}$. Finally, the solid curve is the sum of the contributions to the tuning from $m_{H_u}^2(Q)$ and the threshold corrections, it is very stable across a wide range of $Q$. }}
\label{fig:deltaM3vsQ}
\end{figure}

To derive the threshold correction to $m_{H_u}^2$, we expand around $v_u=v_d=0$,  and read off the coefficient of $|v_u|^2$. Using all the terms, it can be verified analytically that the result is independent of $Q$ up to two-loop order (Equation (5.1) of \cite{Martin:2002iu}). The most relevant terms can be extracted and summarized in a relatively compact formula:
\bea\label{eq:mhusqeff}
& (m_{H_u}^2)_{eff}=m_{H_u}^2 - {3\over 16\pi^2}
y_t^2m_{Q_3}^2(1-\log q_{m_{Q_3}}) + \\
& {y_t^2g_3^2 M_3^2\over 32\pi^4} \Bigg(  (\log x_{m_{Q_3}})^2  - (1-\log q_{m_{Q_3}})^2 + x_{m_{Q_3}}(1-\log q_{m_{Q_3}})
+   2 {\rm Li}_2(1 - x_{m_{Q_3}})\Bigg) \\
&+ (Q_3\to U_3) + \dots
\eea
where the $\dots$ includes other corrections not proportional to $y_t$ at one-loop, and other corrections not involving $M_3$ explicitly at two-loop. Here all of the parameters are running couplings and soft masses evaluated at the scale $Q$, and we have introduced the following notation:
\beq\label{eq:notation}
x_m \equiv {m^2\over M_3^2},\qquad q_m = {m^2\over Q^2}
\eeq

The RG stability of $\Delta_{M_3^2}$ is shown in Fig.~\ref{fig:deltaM3vsQ} for a benchmark point (a similar plot can be made for stops). We see that without including the threshold corrections, the gluino tuning estimate can vary by nearly a factor of 3 when varying the RG scale. With the threshold corrections, it becomes stable to better than 10\%. We also see that the threshold corrections are minimized for $Q$ somewhere between the stop and gluino masses in this example; this makes sense intuitively.

\subsection{Gluino and stop pole mass corrections}

Finally, we will consider the difference between the pole mass  and running mass (item \#5 on the list in the Introduction) for the gluinos and stops. For gluinos, we rely on the classic one-loop results of \cite{Pierce:1996zz}. The one- and two-loop stop thresholds are generally  negligible \cite{Pierce:1996zz,Agashe:1998zz}, even if there is a large splitting between the stop and the first two generations of squarks, but we include them for completeness.

The finite one-loop corrections to the gluino mass are given by gluino and squark loops \cite{Pierce:1996zz}:
\beq\label{eq:gluinopole}
M_3^{\rm pole} = M_3 \left[ 1-\left({\Delta M_3\over M_3}\right)^{g\tilde g}-\left({\Delta M_3\over M_3}\right)^{q\tilde q}\right]^{-1}
\eeq
where 
\beq
\left({\Delta M_3\over M_3}\right)^{g\tilde g} = {3g_3^2\over16\pi^2}(5-3\log q_{M_3} )
\eeq
and 
\beq
\left({\Delta M_3\over M_3}\right)^{q\tilde q} = -{g_3^2\over 32\pi^2}\sum_{i=1}^{12} 
(2-x_{m_{i}}-(1-x_{m_i})^2\log\left| 1-x_{m_i}^{-1}\right| -\log q_{m_i})
\eeq
Here $x_{m_i}, q_{m_i}$ are defined as in \eqref{eq:notation}, and $i$ indexes the 12 squarks  (right-handed and left handed times 6 flavors). 
Again, all the couplings and masses here are running parameters evaluated at the RG scale $Q$. 
The corrections depend on  $Q$ in such a way as to cancel out the $Q$-dependence of the running mass $M_3$ at one-loop order, resulting in a pole mass $M_3^{\rm pole}$ which is largely independent of $Q$ for fixed $M_3(\Lambda)$. 

\begin{figure}[t]
\centering
\includegraphics[width=0.48\textwidth]{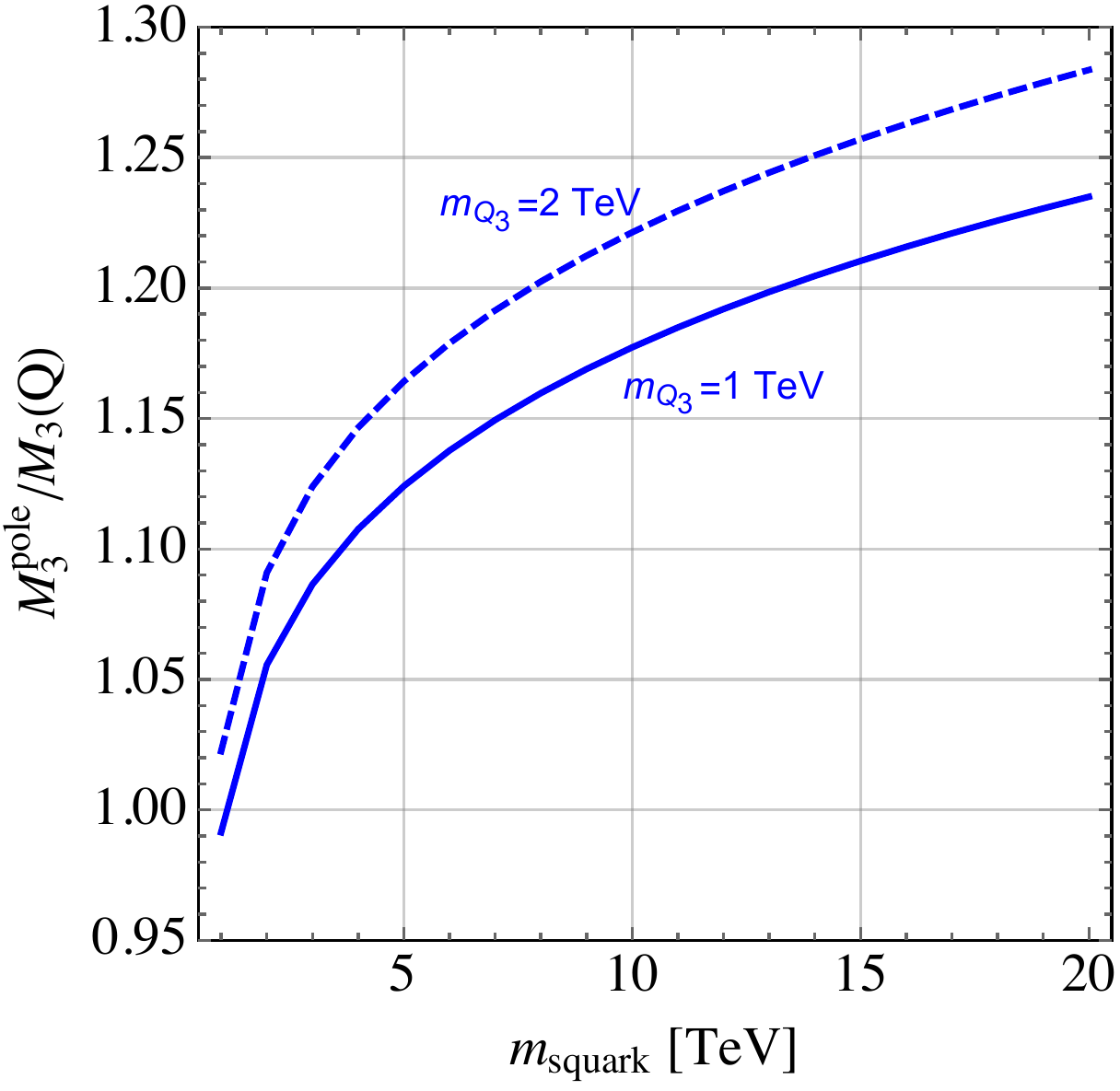}\hfill 
\includegraphics[width=0.48\textwidth]{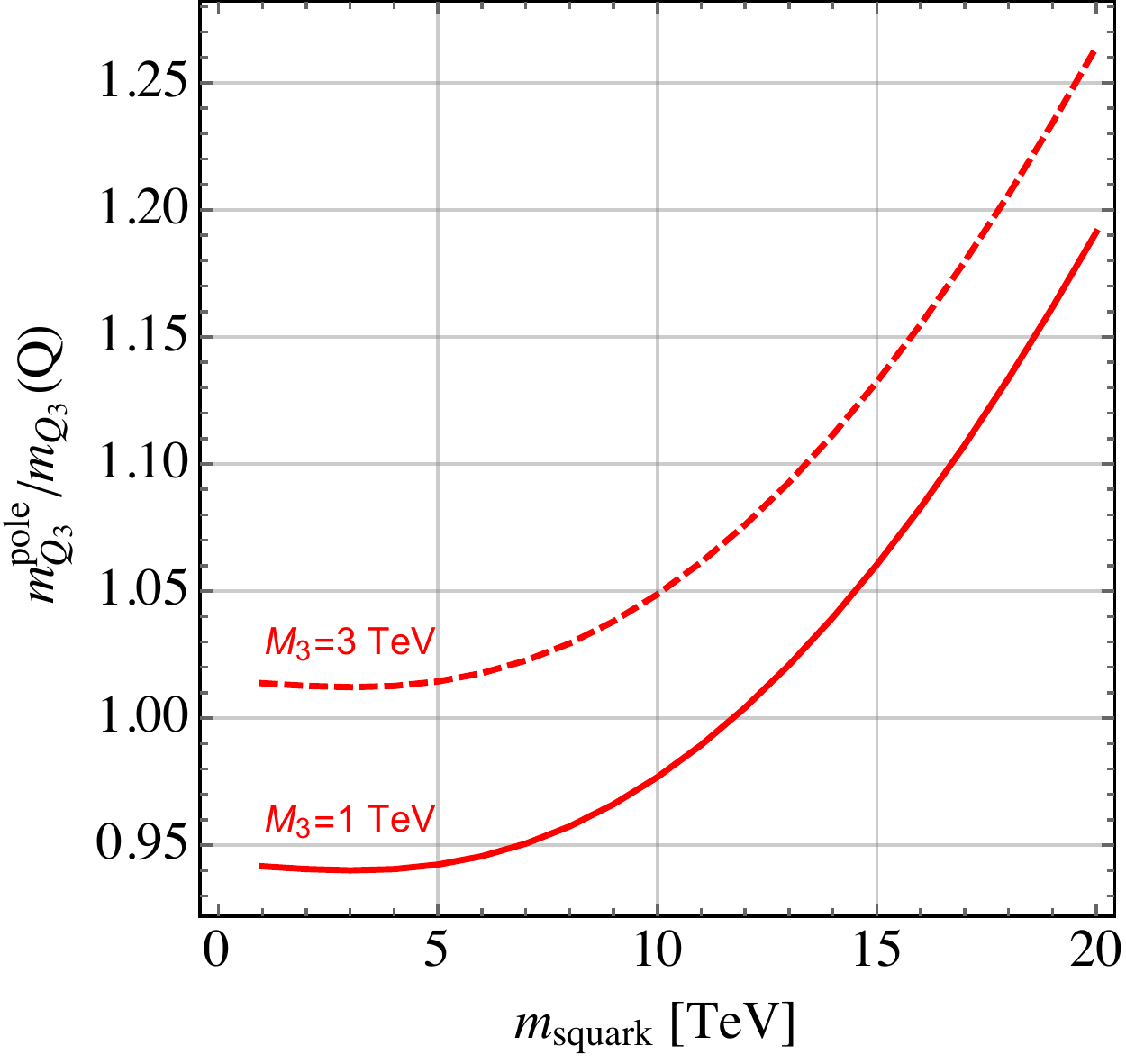}
\caption{\small{Left: Ratio of gluino pole mass to running IR mass (at $Q=1$~TeV), with $M_3(Q)=1.5$~TeV, as a function of  the common 1st/2nd generation squark mass $m_{\rm squark}$, for two different choices of the 3rd generation squark mass. Right: Ratio of the stop pole mass to running IR mass (at $Q=1$~TeV), with $m_{Q_3}(Q)=m_{U_3}(Q)=1.5$~TeV, as a function $m_{\rm squark}$, for two different choices of the gluino mass.}}
\label{fig:polemasses}
\end{figure}

We show the ratio $M_3^{\rm pole}/M_3(Q)$ vs.\ the 1st/2nd generation squark mass (recall our assumption about the squark masses in Section~\ref{sec:xfermatstopgluino}) in Fig.~\ref{fig:polemasses}, left panel, for $Q=1$~TeV and $M_3(Q)=1.5$~TeV. We see that the pole mass is generally even  higher than the IR running mass. At low squark masses, the finite threshold corrections are negligible, but interestingly, at higher squark masses, the finite threshold corrections can be much larger, as high as an additional $\sim 20$\% at $m_{\rm squark}=20$~TeV. This effect would be further amplified if third generation squarks were also made heavy, but we are not interested in this scenario as the stop itself would contribute significantly to the Higgs fine-tuning.

Similarly, the stop pole mass (here $m^2$ stands for either $m_{Q_3}^2$ or $m_{U_3}^2$, and we are neglecting stop mixing) is given by \cite{Pierce:1996zz,Agashe:1998zz}:
\beq\label{eq:stoppole}
(m^2)^{\rm pole}= m^2+ (\Delta m^2)_{\tilde g} 
+ (\Delta m^2)_{\tilde q_{1,2}}
\eeq
where the first term is the IR running squared mass,  the second is the one-loop correction from gluinos,
\bea
(\Delta m^2)_{\tilde g}
&=\frac{g_3^2M_3^2}{6\pi^2}\Big(3+x_m+(x_m+x_m^{-1}-2)\log|x_m-1|\\
&\qquad\qquad \quad +(2-x_m)\log x_m -2\log q_m\Big)
\eea
and the third term is the two-loop contribution from 1st/2nd generations set at a scale $m_{\tilde q_{1,2}}$ \cite{Agashe:1998zz}
\bea\label{eq:stopthr12}
(\Delta m_{Q_{3}}^2)_{\tilde q_{1,2}}=-\frac{m^2_{\tilde q_{1,2}}}{{12\pi^4}}
{g_3^4}
\left( \log 4\pi-\gamma_E +\frac{\pi^2}{3}-2-\log q_{m_{\tilde q_{1,2}}}\right)
\eea
As for the gluino, the threshold corrections included here cancel out the $Q$ dependence of the running mass to a high degree of accuracy. We do not include stop self-corrections proportional to $y_t^2$ which are never more than $\CO(1\%)$ of the running mass, for any reasonable choice of $Q$ and the stop/gluino masses. Similarly, in \eqref{eq:stopthr12} we have omitted terms proportional to $g_{1,2}^4$.

On the right panel of Fig.~\ref{fig:polemasses}, we set $m_{Q_3}(Q)=m_{U_3}(Q)=1.5$~TeV and show the $m_{\rm squark}$ dependence of the ratio of stop pole mass vs.\ running IR mass, again with $Q=1$~TeV. There, we notice the dependence on the gluino mass, with heavier gluinos (dashed) lifting the stop IR mass both through the RGEs and the finite corrections. For larger squark masses, the magnitude of the threshold corrections increases as expected.

\section{Putting it all together}
\label{sec:results}

Having explored several important higher-order effects which impact the calculation of the fine-tuning parameter $\Delta$, we can now combine them and derive more precise naturalness bounds on the physical gluino and squark pole masses as a function of $\Delta$ and $\Lambda$.   As we will see, the combined natural region is not a simple rectangle in the gluino/stop plane. Rather, $\Delta_{M_3^2}$ and $\Delta_{m_{Q_3}^2}$ are nontrivial functions of both the gluino and stop masses. Heavy stops contribute threshold corrections to the gluino pole mass (a relatively minor effect), while gluinos pull up the stops primarily through the RGEs (a much larger effect). As a result, the natural region becomes wedge-shaped.

We will mostly focus on $\Delta=10$ as a benchmark value. (Our full calculation indicates that the limits on the masses for other values of $\Delta$ can be very approximately obtained by rescaling all the masses by $\sqrt{\Delta/10}$.) We will explore the dependence on the 1st/2nd generation squark masses, taking as benchmark values either degenerate squarks, $m_{\tilde q_{1,2}}=m_{\tilde t}$,  or decoupled squarks, $m_{\tilde q_{1,2}}=5$ and 10~TeV.

We have provided all the analytic results necessary to reproduce the plots shown in this section, as well as to explore the parameter space for different benchmark values of $\Delta$, etc.\ should the reader so desire. Specifically, one should calculate the tuning measure $\Delta_{M^2}$ given in (\ref{tuningmeasure1}),
for $M^2=|M_3(\Lambda)|^2$, $m_{Q_3}^2(\Lambda)$ and $m_{U_3}^2(\Lambda)$, using  (\ref{eq:mhusqeff}) for $m_H^2$, and the  transfer matrix relations in sections \ref{sec:xfermatmhu2} and \ref{sec:xfermatstopgluino} to translate the running IR parameters to the UV parameters in order to take the derivative.
Finally, to get a bound on the experimentally accessible masses, one should convert to the gluino and stop pole masses, again using the transfer matrix, and using (\ref{eq:gluinopole}) and (\ref{eq:stoppole}) for the finite threshold corrections.

\subsection{Natural regions in stop/gluino mass plane}

\begin{figure}[t!]
\centering
\includegraphics[width=0.50\columnwidth]{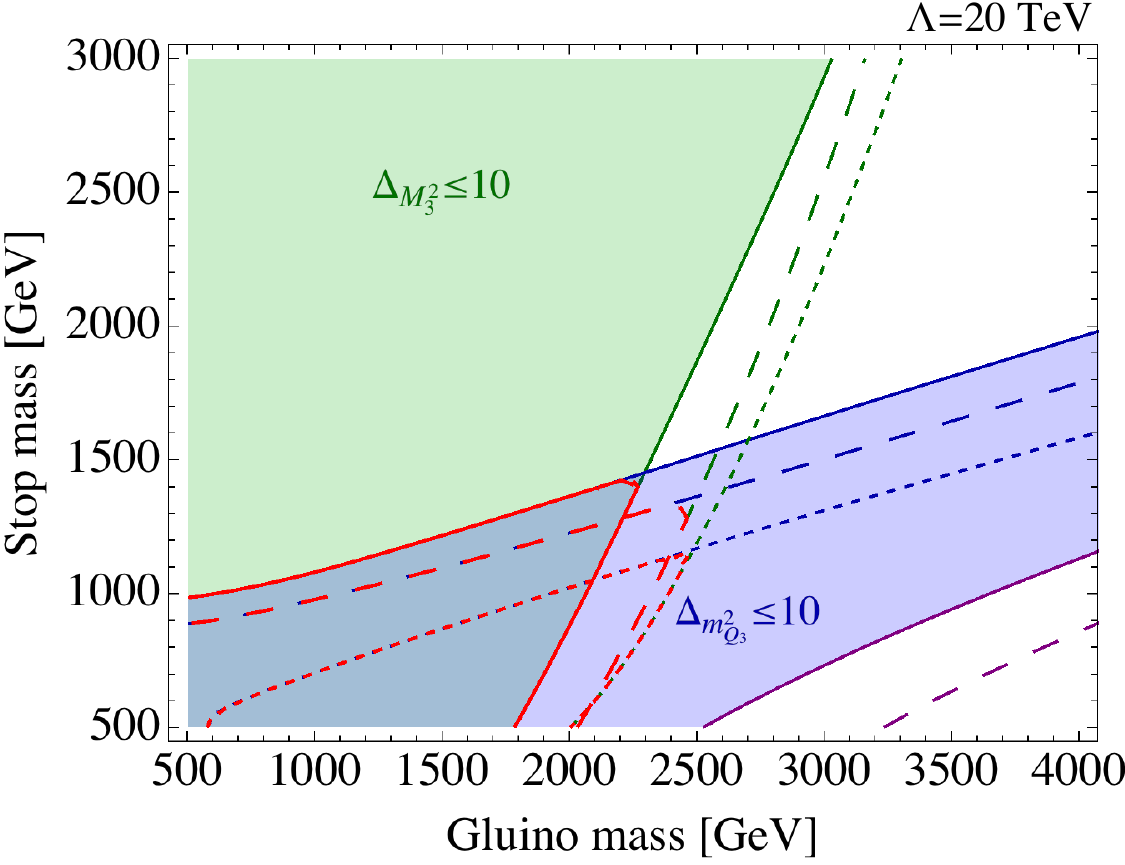}\includegraphics[width=0.50\columnwidth]{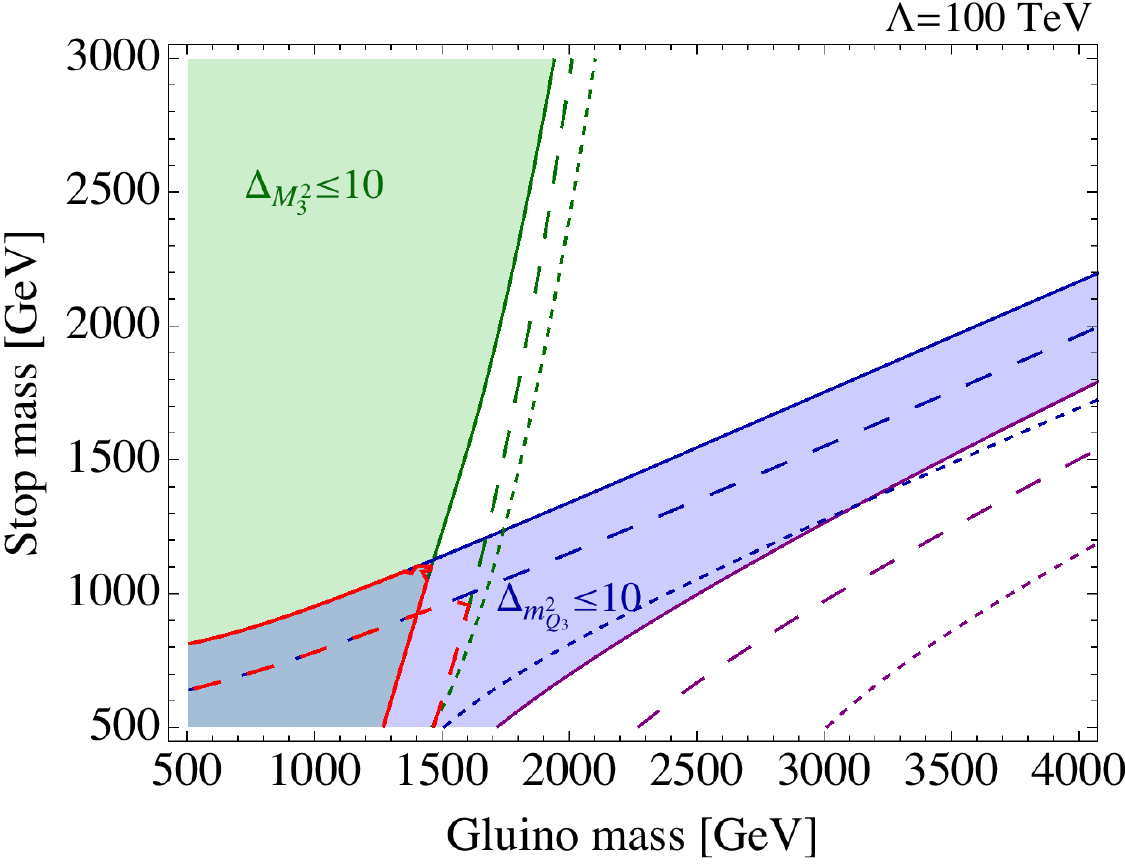}
\includegraphics[width=0.50\columnwidth]{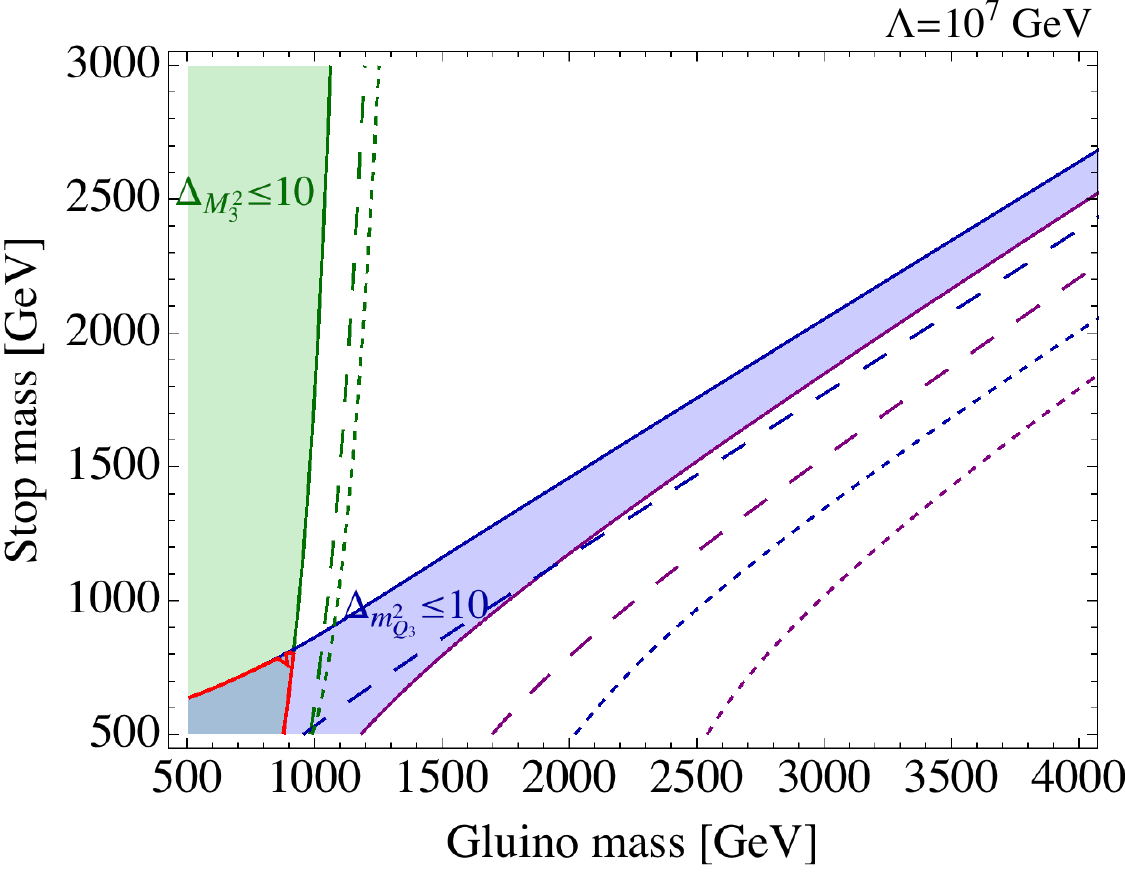}\includegraphics[width=0.50\columnwidth]{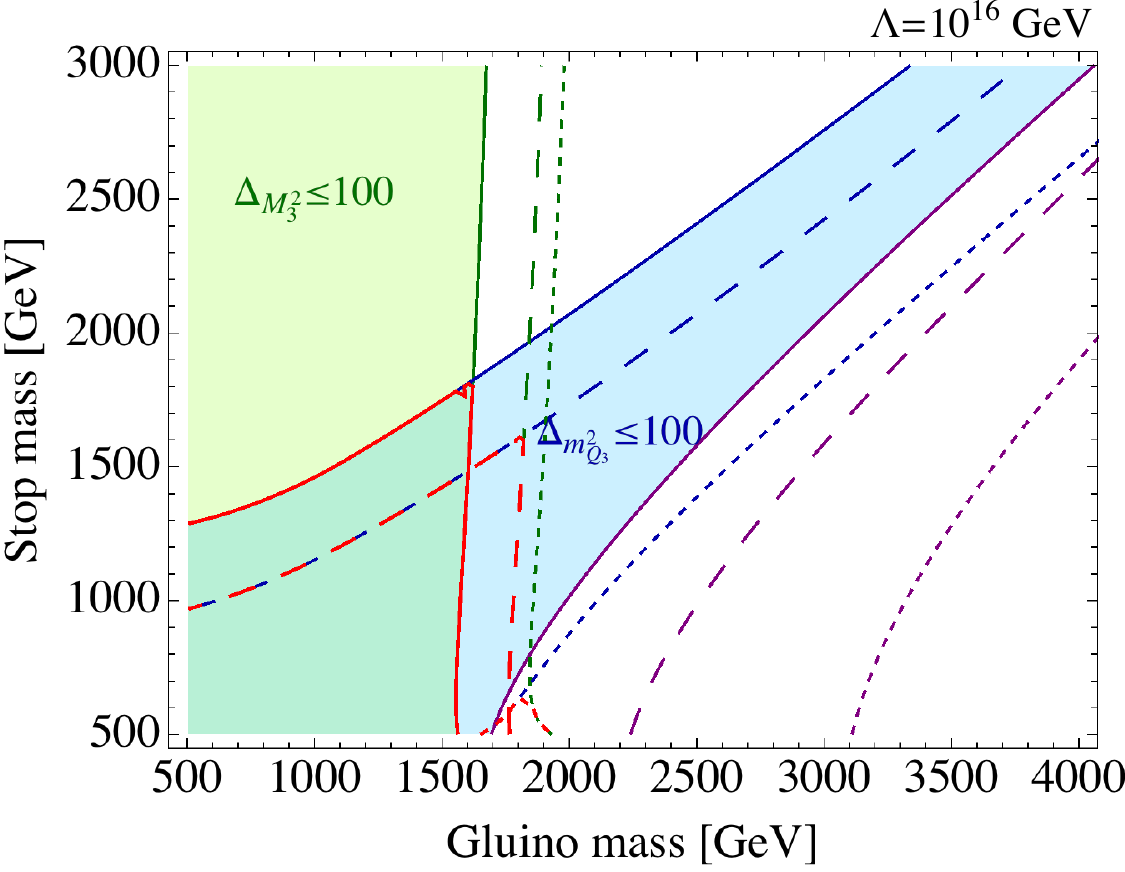}
\caption{\small{ Top: 10\% fine tuned regions with $\Lambda=20$~TeV (left) and $\Lambda=100$~TeV (right), with respect to the gluino mass (green) and the stop mass (blue, delimited by blue and purple lines), for 1st/2nd generation squarks degenerate with the third generation (solid lines), at 5~TeV (dashed lines) and at 10~TeV (dotted lines). The wedge-shaped intersection (delimited by red lines) is the fully natural $\Delta\leq10$ region. 
Bottom: Left:  same as the other plots, but for $\Lambda=10^{7}$~GeV.
Right: same as the other plots, but for $\Lambda=10^{16}$~GeV and 1\% fine-tuning, that is, the lines and shaded regions correspond to $\Delta\leq 100$. }}
\label{fig:wedge}
\end{figure}

Shown in Fig.~\ref{fig:wedge} are the natural regions in the gluino/stop mass plane. As noted above,  for definiteness, we will plot the smaller of the two stop masses according to the maximum of the separate tuning measures. This always corresponds to $\tilde t_L$, due to a larger coefficient in the $m_{H_u}^2$ transfer matrix \eqref{xfermhusq}, from small $SU(2)\times U(1)$ splittings. The tuning bound on $\tilde t_R$ masses is typically less than 5\% higher. We take the messenger scales to be 20~TeV (top, left), 100~TeV (top, right), $10^7$~GeV (bottom, left) and $10^{16}$~GeV (bottom, right). 
The shaded areas mark the regions where the contribution to the Higgs fine-tuning is less than 10\%  (except for $\Lambda=10^{16}$~GeV, where we show 1\% tuned regions instead): in green is the natural region for the gluino, while  in blue we see the stop natural parameter space. The dashed and dotted lines indicate how the natural regions evolve as the 1st/2nd generation squark are taken to 5 and 10~TeV, respectively. Red lines delimit the intersection of gluino and stop regions, i.e.\ the region in which both gluinos and stops are natural.

Although it does not impact our 10\% natural region, it is interesting that the stop natural region is actually a {\it strip} (with the upper and lower bounds delimited by blue and purple lines), with the lower boundary corresponding to large {\it and negative} UV stop mass squared, which have been pulled up by the gluino to be non-tachyonic in the IR. The slope of the band increases with $\Lambda$, due to the increased dependence on the gluino mass from RG running.

\begin{figure}[t!]
\begin{tabular}{ll}
\includegraphics[width=0.46\textwidth]{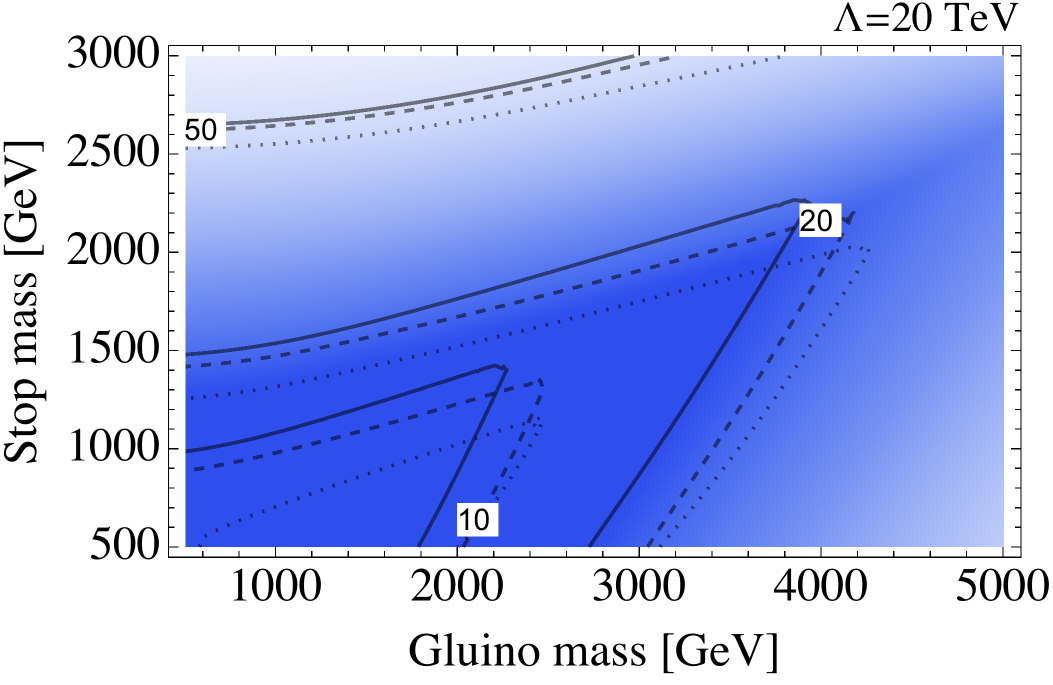}\includegraphics[width=0.46\columnwidth]{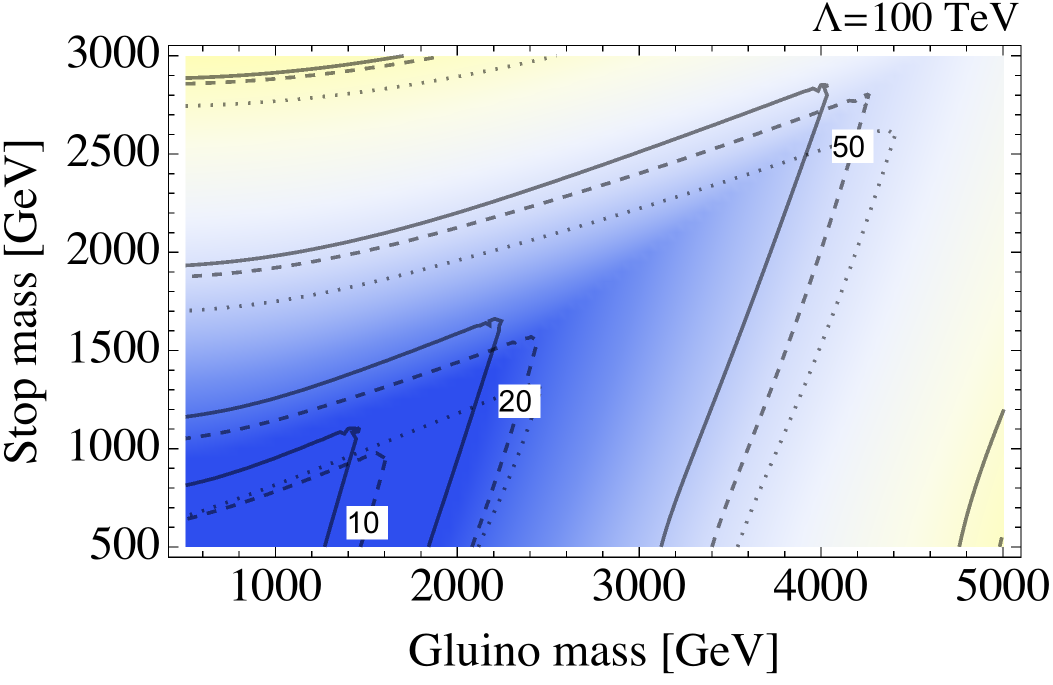} & 
\multirow{2}{*}[3.3cm]{\includegraphics[scale=0.8]{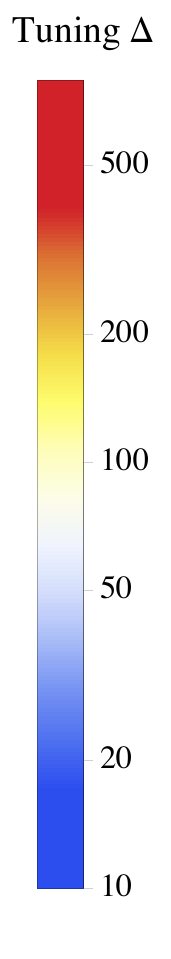}
}
\\
\includegraphics[width=0.46\columnwidth]{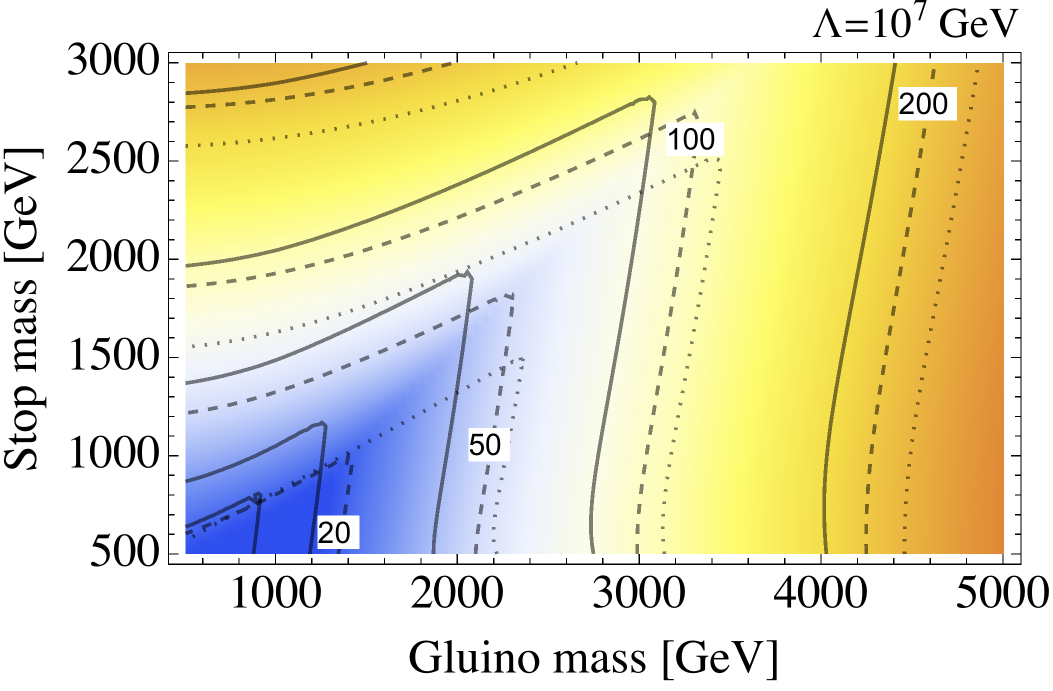}\includegraphics[width=0.46\columnwidth]{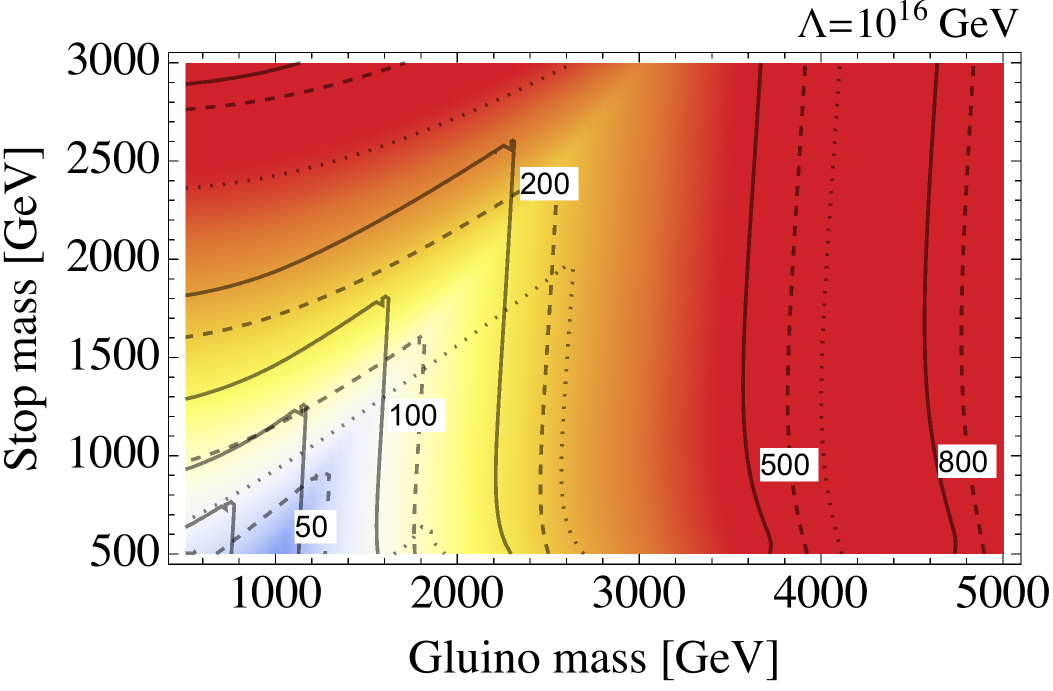}
&
\end{tabular}
\caption{Level of Higgs fine-tuning in the gluino-stop mass plane, for different values of the messenger scale $\Lambda=20$~TeV (top, left), $100$~TeV (top, right), $10^7$~GeV (bottom, left) and $10^{16}$~GeV (bottom, right). Different values of fine-tuning are color-coded according to the legend on the right, with shades of blue corresponding to fine-tuning levels close to 10\%, whites in the few percent range, and yellows/reds for sub-percent fine-tuning (for definiteness, the color-coding corresponds to 1st and 2nd generation squarks at 5~TeV). Contours for specific values are also provided, with solid, dashed and dotted contours corresponding to squarks degenerate with the stops and 5 and 10~TeV.}
\label{fig:tuning_all}
\end{figure}

It can also be seen that raising the 1st/2nd generation squark masses can expand the maximum natural gluino mass through the 1-loop threshold correction, but it reduces the maximum natural stop  mass, through the 2-loop RGE and threshold corrections. (Perhaps in certain extensions of the MSSM, this 2-loop effect could be alleviated \cite{Hisano:2000wy}; this would be interesting to explore in future work.) This trade-off becomes worse at higher messenger scales (longer running): for example, for $\Lambda\ge 100$~TeV and $m_{\tilde q_{1,2}}=10$~TeV, there is simply  no $\Delta\le 10$ fully natural region for both gluinos and stops, as the dotted lines do not intersect. For squarks at 5~TeV, the same happens at $\Lambda\ge 10^7$~GeV (now the dashed lines do not intersect). With the experimental limits on degenerate squarks presented in \cite{Buckley:2016kvr} at around 2~TeV for the $R$-parity conserving MSSM, it can be seen that 1st and 2nd generation squarks at 2--5~TeV occupy a ``sweet spot'': heavy enough to not be efficiently produced at the LHC, but light enough to not contribute too negatively to the stop tuning. Further raising the squark mass does not significantly improve improve the gluino tuning (via the threshold corrections), but considerably lowers the allowed stop mass.

In Fig.~\ref{fig:tuning_all}, we show the level of fine-tuning in the gluino-stop mass plane, with varying 1st and 2nd generation masses shown as solid, dashed and dotted lines.
We provide these figures as a reference on which future LHC limits can easily be superimposed to assess the fine-tuning of SUSY. It is easily noted that even relatively mild level of tuning such as 5\% will hardly be probed at the LHC for messenger scales below 100~TeV; in this sense, our reference choice of 10\% fine-tuning also represents a target that the LHC can comprehensively exclude \cite{Buckley:2016kvr}. 

\subsection{Absolute upper limits on gluinos and stops}

\begin{figure}[t!]
\centering
\includegraphics[width=0.5\columnwidth]{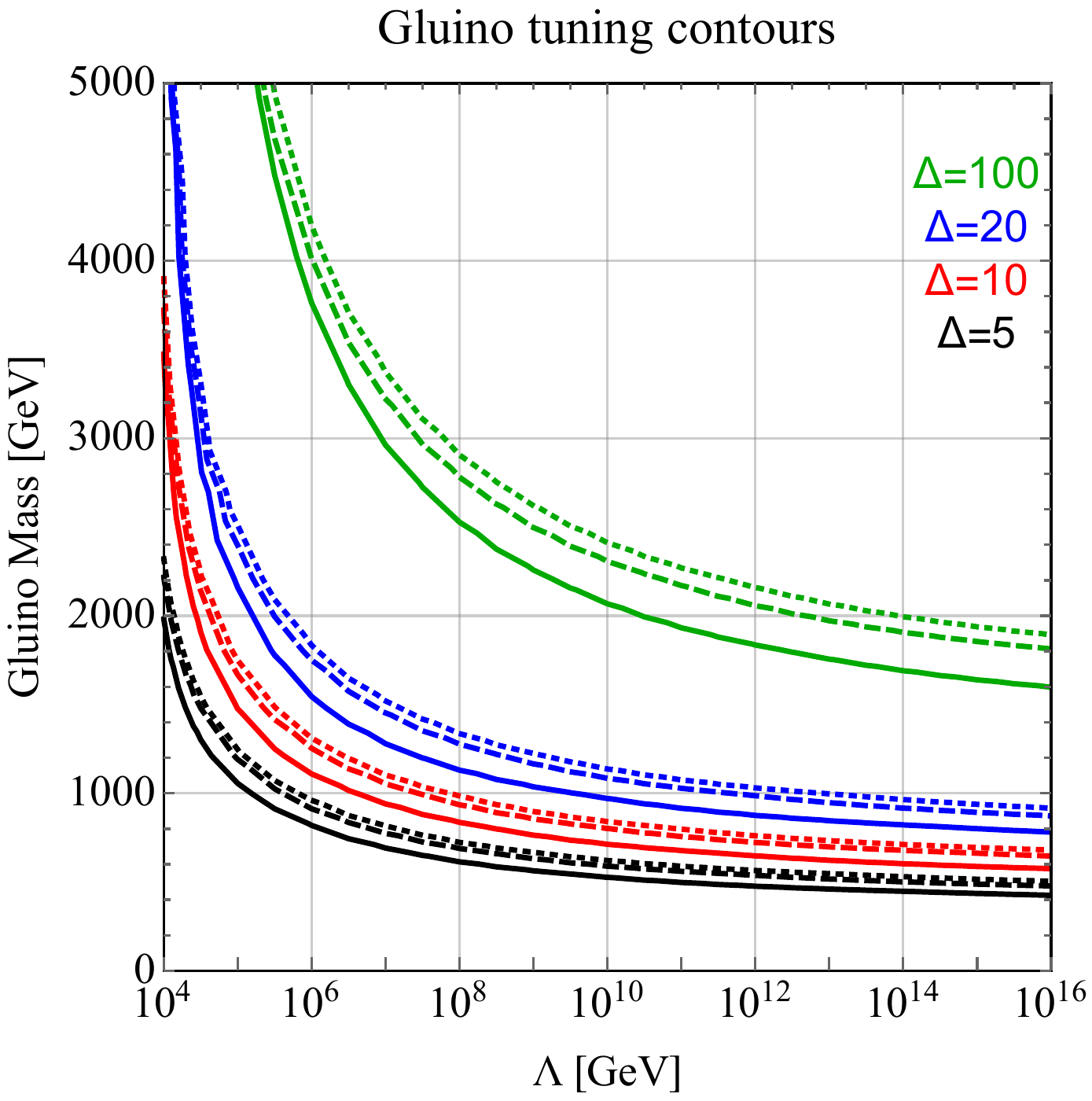}\includegraphics[width=0.5\columnwidth]{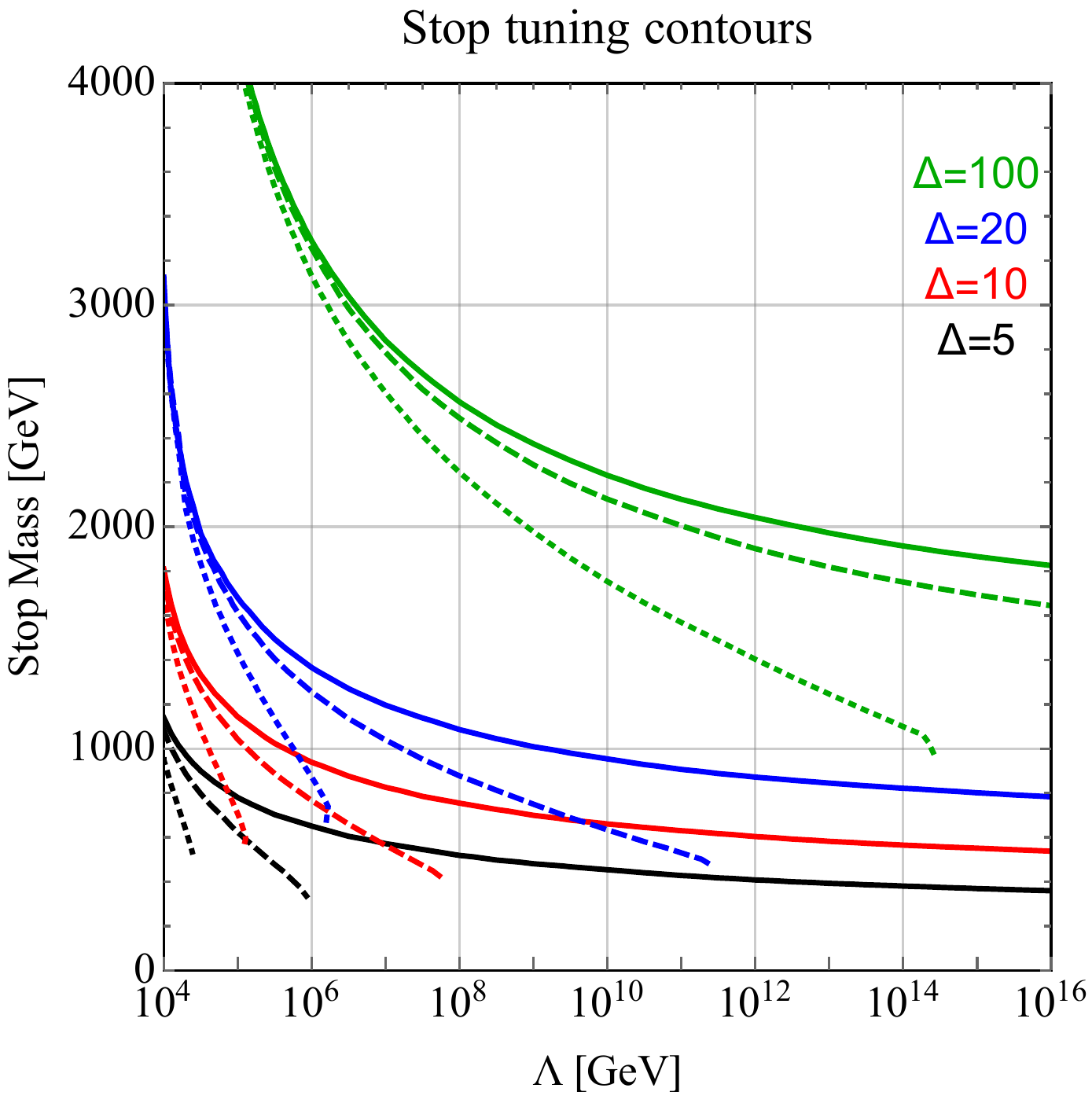}
\caption{\small{Tuning contours of IR gluino (left) and stop mass (right) for $\Delta=5$ (black), 10 (red), 20 (blue) and 100 (green) as a function of messenger scale $\Lambda$. In both plots, the 1st and 2nd generation squark masses are set equal to the 3rd generation mass (solid lines), at 5~TeV (dashed lines), and 10~TeV (dotted lines).}
}
\label{fig:tuning}
\end{figure}

It is also informative to project our two-dimensional tuning regions onto the gluino or stop axes and obtain the absolute upper bounds on the physical gluino and stop masses, as a function of the messenger scale $\Lambda$, for different values of $\Delta$. (This amounts to taking the tips of the wedge regions in fig.~\ref{fig:wedge}, but across a wider range of messenger scales.)
The result of this projection is shown in 
the left and right panel of Fig.~\ref{fig:tuning}.  In Table~\ref{tab:tuning}, we also list some reference values for the physical gluino and stop masses corresponding to $\Delta=10$, for different choices of $\Lambda$ and the 1st/2nd generation squark masses. 
As in fig.~\ref{fig:wedge}, the solid, dashed and dotted lines correspond to 1st/2nd generation squark masses degenerate with the 3rd generation, or at 5~TeV and 10~TeV respectively. These plots illustrate the strong dependence on the messenger scale -- the tuning bounds decrease sharply from $\Lambda=10$~TeV to $\Lambda\sim 10^3$~TeV. The plots also demonstrate the danger of increasing the 1st/2nd generation squark masses -- the stops go tachyonic very quickly for larger messenger scales, and even when they are not tachyonic, the tuning bounds become increasingly restrictive. 

\begin{table}[t]
\resizebox{\columnwidth}{!}{
\begin{tabular}{|c|c|c|c|c|c|c|} \hline
 & \multicolumn{2}{c|}{$m_{\tilde{q}} = m_{\tilde{t}}$} & \multicolumn{2}{c|}{$m_{\tilde{q}} = 5$~TeV} & \multicolumn{2}{c|}{$m_{\tilde{q}} = 10$~TeV} \\ \hline
 & $\Lambda = 20$~TeV & $\Lambda = 100$~TeV & $\Lambda = 20$~TeV & $\Lambda = 100$~TeV & $\Lambda = 20$~TeV & $\Lambda = 100$~TeV \\ \hline \hline
$m_{\tilde{g}}$~(GeV) & 2230  & 1475 & 2480 & 1665 & 2600 & 1750 \\ \hline
$m_{\tilde{t}}$~(GeV) & 1455 & 1140 & 1400 & 1040 & 1260 & 705 \\ \hline
\end{tabular}
\caption{Upper limits on the mass of the gluino $m_{\tilde{g}}$ and stop squark $m_{\tilde{t}}$ requiring $\Delta = 10$ for varying values of 1st and 2nd generation squarks $m_{\tilde{q}}$ and messenger scale $\Lambda$ (see Fig.~\ref{fig:tuning}).} 
\label{tab:tuning}
}
\end{table}

\subsection{Comparison with LL approximation}
\begin{figure}[t]
\centering
\includegraphics[width=0.5\columnwidth]{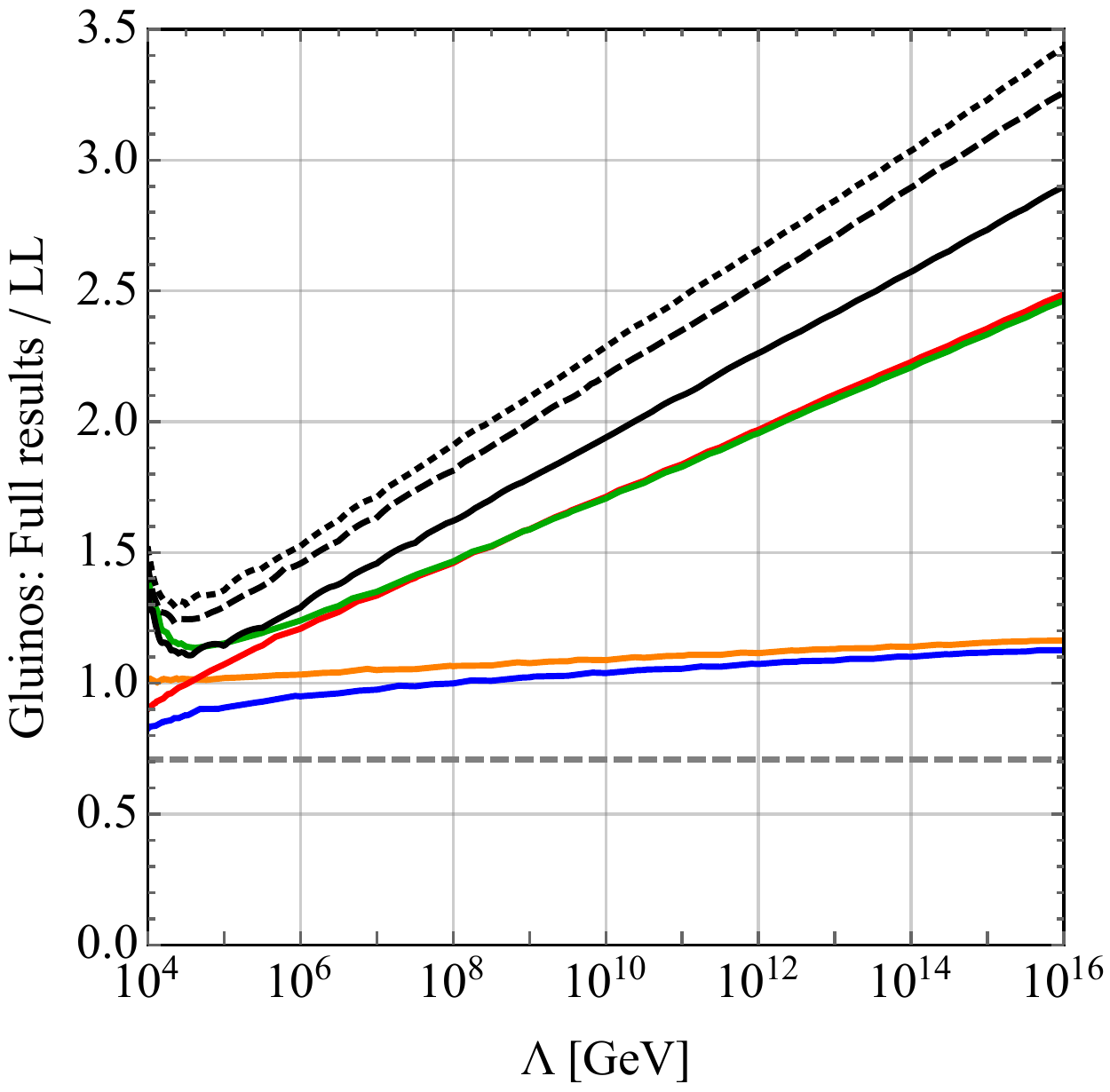}\includegraphics[width=0.5\columnwidth]{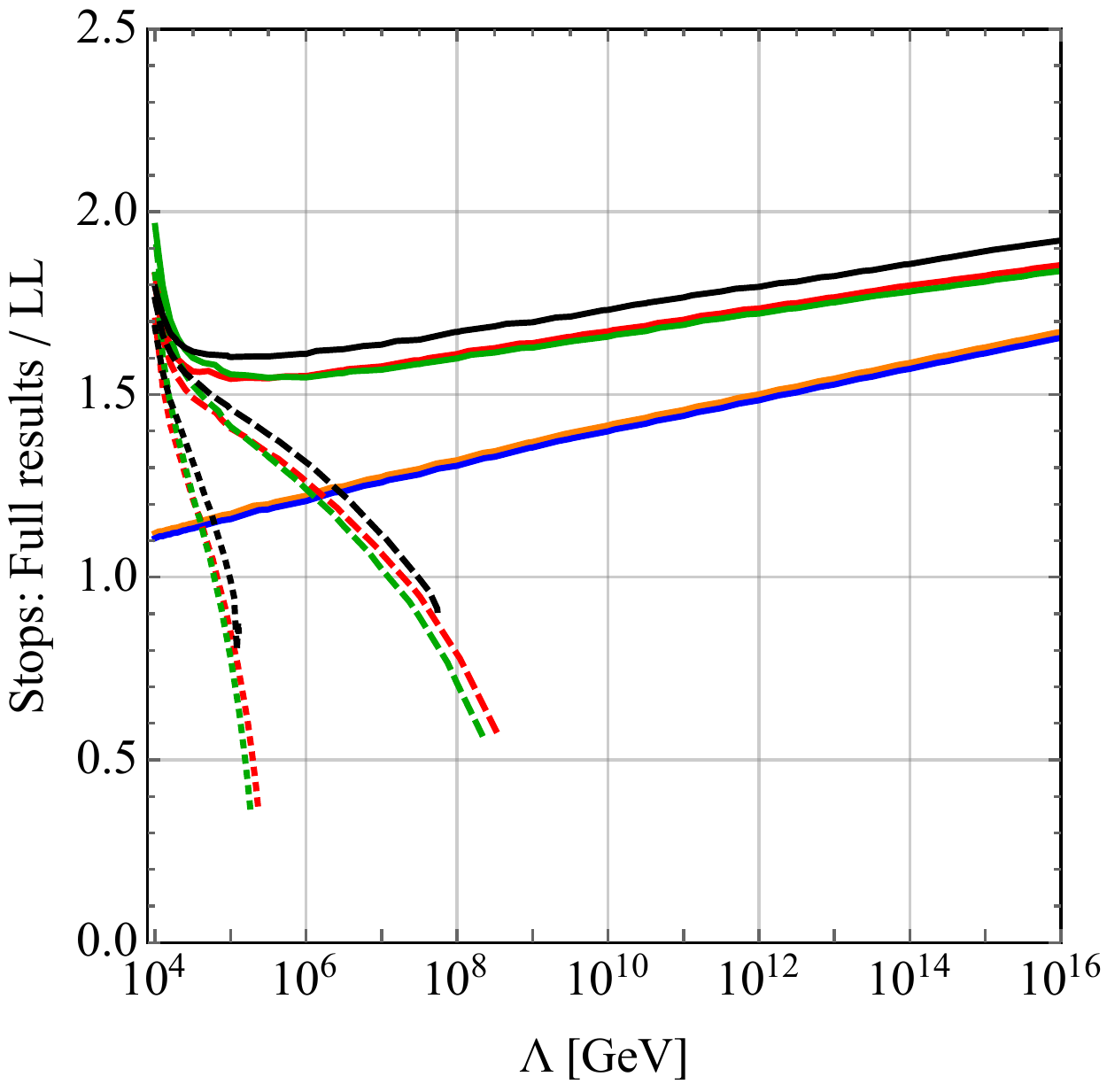}
\caption{\small{The ratio of the $\Delta=10$ naturalness bound on the gluino (left) and stop (right) masses, with the higher-order effects outlined in this paper sequentially added, to the same naturalness bound in the LL approximation (with the couplings evaluated at $Q=1$~TeV). 
The successively included effects are: resummed one-loop RGEs (orange), resummed two-loop RGEs (blue), IR running masses (red), two-loop threshold corrections to $m_{H_u}^2$ (green), and finally moving converting the IR running mass to the pole mass (black). First and second generation squarks are varied between being degenerate with 3rd generation (solid), or at 5~TeV (dashed) and 10~TeV (dotted). The dashed horizontal line for the gluino is the LL result of \cite{Papucci:2011wy}, with a numerical error resulting in a $\sqrt{2}$ reduction.  
}}
\label{fig:tuning_compare}
\end{figure}

Finally, we exhibit the effect on the naturalness bounds of adding each correction considered in Section~\ref{sec:tuning} in turn. Shown in Fig.~\ref{fig:tuning_compare}  is 
the ratio of the corrected naturalness bound on the gluino and stop masses relative to the leading order calculation from \eqref{deltamhusqstop}--\eqref{deltamhusqgluino}, as each higher-order effect is added. For the LL estimate, we have taken $y_t$ and $g_3$ at $Q=1$~TeV.  As in previous plots, we take the UV gluino and stop masses saturating their 10\% naturalness bounds. The final result is a gluino mass which is at least 10--30\% larger than the corresponding leading order calculation for the same value of $\Delta$, and a stop mass at least 50\% larger when squarks are light. For the gluino, the dominant effects are: the difference between IR and UV masses (high messenger scales) and the threshold corrections to $m_{H_u}^2$ (low messenger scales). Also important are the gluino pole mass corrections from the heavy 1st/2nd generation squarks. Meanwhile, for the stop, the dominant factor is the difference between IR and UV masses (especially the additive boost from the gluinos and the drop due to 1st/2nd generation squarks), with the other effects changing the allowed stop mass just by a few percent. 

\section{Conclusions}
\label{sec:conclusion}

In this work, we have detailed several precision corrections to the fine-tuning of the Higgs mass. With SUSY increasingly under pressure from the second run of the LHC, our accurate estimates in Fig.~\ref{fig:wedge} and Fig.~\ref{fig:tuning_all} of what constitutes a fully-natural SUSY spectrum can be used as points of reference as more data is collected. In \cite{Buckley:2016kvr}, we have explored the collider consequences of the natural spectra described here, and found that only very low messenger scales, $\Lambda\lesssim 100$~TeV are compatible with 10\% fine-tuned SUSY after the first $\sim$15 fb$^{-1}$ of 13 TeV LHC data. 

Motivated by the latest LHC constraints, we have given special attention in this work to ``Effective SUSY" scenarios where the 1st/2nd generation squarks are heavier than the 3rd generation. We have uncovered significant new corrections to the tuning bounds in this scenario. While increasing the 1st/2nd generation squark masses moderately relaxes the gluino tuning bound through one-loop finite threshold corrections, it significantly strengthens the stop tuning bound through the two-loop RGEs. The tension between these (together with the LHC constraints) leads to a ``sweet spot" of $m_{\tilde q_{1,2}}\sim2$--5~TeV. One very interesting future direction for model-building will be to investigate viable UV completions of this moderate Effective SUSY scenario consistent with the low messenger scales $\Lambda\lesssim 100$~TeV required by the current LHC constraints, see \cite{ArkaniHamed:1997fq,Gabella:2007cp,Aharony:2010ch,Craig:2011yk,Hardy:2013uxa} for some promising models and  \cite{Buckley:2016kvr} for further discussion of this.

Given the strong tension between the tuning bounds derived here and the current LHC constraints, another interesting direction would be to challenge the underlying assumptions going into the tuning calculations. In general, any extension or modification of the MSSM between the messenger scale and the weak scale has the potential of significantly changing our tuning calculation, and a similarly precise calculation should be carried out for that case. For example by introducing ``super-safe'' Dirac gluinos \cite{Fox:2002bu,Kribs:2012gx}, one could put the gluinos out of reach of current bounds without incurring as much of a fine-tuning price. However, Dirac gluinos would also change the RGEs for the stop mass-squared, removing the dependence on the gluino mass in the running \cite{Fox:2002bu}. In this way, the gluino tuning is ameliorated with respect to the MSSM, but the stop tuning is actually worsened, and experimental stop limits would be more constraining. As the SUSY production rates also change significantly (in particular, there is no gluino $t$-channel diagram giving large valence squark production), it would be very interesting to revisit Dirac gluinos with an eye towards precision corrections to the tuning calculation, combined with recasted limits on simplified models.

Another possibility is to relax the tuning bound on the stop mass through the addition of new particles which positively affect the stop RGE (e.g.~the addition of vector-like quarks as in \cite{Hisano:2000wy}). In the MSSM, we found that heavy 1st and 2nd generation squarks push down the mass of a ``natural'' stop through the RG equations and in the threshold corrections. New particles could potentially counteract this, allowing  $\Delta = 10$ tuning with heavier stops than considered in this paper. Similar conclusions hold for the NMSSM, as described in \cite{Hall:2011aa} at the LL level, where an extra singlet lifts the tree-level Higgs mass and stop contributions to fine-tuning are reduced due to sizable mixing between the singlet and $H_u$ (although this should be revisited in light of the Higgs couplings being rather SM-like, see e.g. \cite{Agashe:2012zq,Gherghetta:2012gb}).

We have not considered the role of the higgsino mass $\mu$ in this work, because its precision corrections are rather small in the MSSM. However, models where the higgsino mass is not set primarily by the $\mu$ term, e.g.~\cite{Dimopoulos:2014aua,Martin:2015eca,Cohen:2015ala,Garcia:2015sfa,Delgado:2016vib}, are a very interesting loophole to the tuning bounds and a promising direction for future work. Not only would allowing for heavier higgsinos have a potentially huge effect on the collider phenomenology discussed in \cite{Buckley:2016kvr}, but models of this type may be sufficiently removed from the MSSM that the tuning calculations for stops and gluinos would also be significantly impacted.

Finally, a note on our assumptions about fine-tuning: we have here shown the natural regions given our measure \eqref{tuningmeasure1} and taking the maximum when multiple sources for the Higgs tuning are present. If the UV parameters are assumed to be independent, one might hope to take into account multiple tunings by adding them together somehow, e.g.~in quadrature. In this case the wedges in Fig.~\ref{fig:wedge} would be rounded at the tips and would reduce the maximum allowed masses for gluinos and stops by up to about 200~GeV, leaving our results qualitatively unchanged. A separate aspect is that the mass of the stop (a scalar) could  also be susceptible to tuning: for a light stop and much heavier gluino, the stop itself suffers from a naturalness problem \cite{Arvanitaki:2013yja}. This tends to be an issue only for high messenger scales or extremely light stops  -- neither of which is well-motivated given the latest LHC bounds \cite{Buckley:2016kvr}.

\section*{Acknowledgements}

We are grateful to J.~A.~Casas, J.~Moreno, K.~Rolbiecki, S.~Robles, J.~Ruderman, S.~Thomas and B.~Zaldivar for helpful discussions.
The work of AM and DS are supported by DOE grant DE-SC0013678.

\bibliographystyle{utphys}
\bibliography{tuning_natural}

\end{document}